\documentclass[amsmath,amssym,ams, preprint, superscriptaddress,nofootinbib]{article} 
\usepackage[a4paper,textwidth=17.9cm,textheight=26cm]{geometry}
\usepackage[english]{babel}
\usepackage[explicit]{titlesec}
 
\usepackage[T1]{fontenc}
\usepackage{lmodern}
\usepackage{helvet}
\usepackage{lineno}
\usepackage{graphicx}
\usepackage{xr}
\usepackage{authblk}
\usepackage{ulem}
\usepackage{siunitx}
\usepackage{upgreek}
\usepackage{amsmath}

\newcommand{\nocontentsline}[3]{}
\newcommand{\tocless}[2]{\bgroup\let\addcontentsline=\nocontentsline#1{#2}\egroup}

\makeatletter
\newcommand*{\addFileDependency}[1]{
  \typeout{(#1)}
  \@addtofilelist{#1}
  \IfFileExists{#1}{}{\typeout{No file #1.}}
}
\makeatother

\addto\captionsenglish{}
\addto\captionsenglish{}

\begin{document}

\title{Metasurface-controlled high-speed tunable external cavity lasers}
\author[1, 2]{Zahra Basiri}
\author[1, 2]{Alessandro Tomasino}
\author[1, 2]{Gabriel Jülg}

\author[3]{Andrea Lanfranchi}
\author[1, 2]{Ileana-Cristina Benea-Chelmus}

\affil[1]{Hybrid Photonics Laboratory, École Polytechnique Fédérale de Lausanne (EPFL),  CH-1015, Switzerland}
\affil[2]{Center for Quantum Science and Engineering (EPFL), CH-1015, Switzerland}
\affil[3]{Department of Chemistry and Industrial Chemistry, University of
Genoa, 16146, Genova, Italy}

\date{\today}
\maketitle


\begin{abstract}
Tunable lasers are essential for optical communication, spectroscopy, and precision sensing, where flexible and fast control of the laser wavelength is needed. However, conventional tunable laser systems often rely on mechanical actuation, which limits their tuning speed, stability, and repeatability. Alternative tuning methods, such as adjusting the temperature of the gain medium or the injection current, are either slow or suffer from unwanted coupling between frequency and intensity. Electro-optic modulators implemented after the laser offer fast frequency tunability but these components are bulky and prone to high insertion losses. Here, we introduce an external cavity semiconductor laser that achieves ultrafast frequency and amplitude tunability within a compact footprint without any moving parts. The laser employs a resonant electro-optic metasurface as the external mirror, providing narrowband feedback that enables simultaneous single-mode operation and voltage-controlled tuning across a wide range of currents.  Leveraging a narrow linewidth design, we demonstrate mode-hop-free linear frequency tuning with rates of up to 70 THz/s, and frequency excursions reaching 110 MHz under a 10 V peak RF drive, clearly surpassing rates attainable through mechanical actuation. Moreover, by biasing the laser near its threshold, our platform enables intensity modulation  with an efficiency above 100$\%$ at a peak voltage of 4~V. These results highlight the platform’s potential for applications requiring high-speed and mode-hop-free tunable lasers, such as free-space optical communication, laser ranging, and high-resolution spectroscopy.
\end{abstract}

\tocless\section{Introduction}\label{sec_intro}
Compact and efficient control of laser emission is an essential feature in diverse scientific and technological contexts, including laser ranging~\cite{liang2024evolution,zhang2022large}, high-resolution spectroscopy~\cite{giorgetta2010fast,letokhov1975nonlinear}, optical communications~\cite{willner2021orbital,liu2019vertical,wang2012terabit}, and 3D printing~\cite{jonuvsauskas2019mesoscale}. Since the mid-1970s~\cite{voumard1977external}, tremendous efforts have gone into the development of external cavity diode laser (ECL) designs with a broad range of configurations optimized for wavelength tunability and spectral purity~\cite{harvey1991external,lang1980external,heim2025hybrid,lihachev2022low,li2022integrated}. Early proposals relied predominantly on mechanical tuning, which limited speed and stability. To enable faster, voltage-controlled actuation, electro-optic (EO) crystals such as lithium niobate~\cite{bLN1994} and lithium tantalate~\cite{LT002} were later introduced as tuning elements~\cite{ADP}. However, modulators made from these materials were bulky and required a grating in addition to the modulator to simultaneously achieve single-mode operation and high-speed control over the laser wavelength~\cite{ye2007tunable}. These multi-component systems were prone to multiple reflections and misalignment over time, ultimately impacting the long-term stability of the entire system. 

The lack of compact modulators put a roadblock to reducing the system footprint while maintaining fast and single mode operation. Recent advances in integrated photonic platforms have begun to propose solutions by utilizing on-chip EO and piezoelectric modulators that enable ultrafast tunable lasers~\cite{heim2025hybrid,lihachev2022low,li2022integrated,liang2015ultralow}. Yet, similar capabilities have not been achieved for free-space laser systems, where the intra-cavity light propagates in free space rather than in a waveguide~\cite{ossiander2023metasurface}. In contrast to integrated photonic lasers, these systems would be immediately applicable to free-space applications such as imaging~\cite{wu2017ultrafast,kim2024metasurface,reshef2021optic}, ranging~\cite{zou2022high,kim2021nanophotonics,park2021all,industrial}, diffractive optics, and light projection~\cite{dong2025motion,yang2023ultrahigh,butow2024generating}. They are also attractive for applications where propagation in fibers or along waveguides can lead to uncontrolled dispersion, nonlinearities, and temperature fluctuations~\cite{yang2022noise,rizzelli2022phase,cartaxo2002influence}. 

Optical metasurfaces, i.e. planar arrays of subwavelength elements, condense the functionalities of multiple bulky and complex optical components into a single, engineered two-dimensional surface~\cite{shaltout2019spatiotemporal,camera2,retro,spectrometer,chen2020flat,rubin2019matrix,aieta2015multiwavelength,yu2012broadband,kildishev2013planar,meng2025ultranarrow,li2018addressable,memssteer,ossiander2023metasurface}. Their compactness makes them particularly important in applications requiring a small footprint, along with improved long-term robustness and stability. In most applications, metasurfaces are used downstream of the laser, where they manipulate the space-time characteristics of the light with minimal component overhead, simplifying an optical setup~\cite{camera2}. In this case, they do not affect the lasing process.

More intriguingly, metasurfaces can be implemented directly as part of a laser cavity, where their multifunctionality enables custom-tailored and dynamic feedback to the gain medium. This integration allows control over the spatial, spectral, and polarization properties of laser emission~\cite{structuredlaser}. Two main strategies have emerged for the concrete cavity design. One approach relies on patterning sub-wavelength structures into the gain medium, effectively transforming it into a metasurface. This has been used to statically shape the beam profile of perovskite lasers~\cite{ultrafast}, implement low threshold semiconductor lasers~\cite{zhong2023ultra,hwang2021ultralow} or control the polarization state of terahertz quantum-cascade lasers~\cite{thzqcl,curwen2019broadband,thzvecsel,mtseclthz}. A second approach uses metasurfaces as engineered mirrors of the laser cavity, either patterned directly on the laser's facet or in an external cavity configuration. While static metasurfaces have been patterned directly onto the laser facet to achieve spin-dependent beam shaping~\cite{jones} or mode-selective metasurface vertical cavity surface emitting lasers~\cite{modeslectivevecsel}, or designing spatially tunable photonic-crystal surface-emitting lasers~\cite{luan2023reconfigurable,yang2020spin,yoshida2023high}, incorporating a tunable mechanism into these designs remains a tremendous task due to space limitations and fabrication constraints. To address this limitation, external cavities have been proposed. In this case, the reflection is suppressed at one of the laser's facets with an anti-reflection coating and tailoring the external mirror provides flexibility in controlling laser emission. Following this approach, a multifunctional metasurface external cavity laser~\cite{ossiander2023metasurface} has been proposed, but the lack of electrical tunability necessitated mechanical adjustments for wavelength control. 

Meanwhile, several approaches have been explored to make metasurfaces electrically tunable. Among active tuning methods, electro-optic metasurfaces offer fast, voltage-driven control in a compact form~\cite{spatioactive,riseofeo,park2021all,gu2023reconfigurable,industrial,wu2019dynamic}. They often rely on resonant metasurface designs, which exploit narrow resonances as a means to provide reflection or transmission only within a desired frequency band. This makes them uniquely suited for achieving single-mode laser emission. Furthermore, the radiative nature of these resonances can often be controlled through geometry, enabling reflection characteristics with custom-tailored bandwidth.  Further integrating Pockels materials within such metasurfaces enables electrical control over these characteristics~\cite{riseofeo,weigand2021enhanced,LNO2,LNO3,jrd1ultrahigh,beneaelectro} at speeds up to gigahertz frequencies~\cite{beneagigahertz}, opening up the intriguing question whether these tunable metasurfaces can facilitate ultrafast tunable lasers.

Here, we demonstrate the integration of an ultrathin electro-optic modulator with a laser medium to simultaneously engineer the optical spectrum and enable high-speed modulation of laser light in an external cavity configuration. By exploiting guided mode resonances, we provide frequency-dependent feedback to a semiconductor laser diode operating in the telecom C-band and achieve single-mode operation. The guided mode resonances form within a single layer of electro-optic material onto which we pattern an array of metallic electrodes, similar to our previous work~\cite{beneaelectro}. We show that tweaking the array periodicity enables lasing at engineered frequencies across the gain bandwidth, while matching the gain with the resonance frequency of the guided modes influences laser threshold. Enabled by the large electro-optic coefficient of the organic layer, we provide voltage-tunable feedback to the gain medium, enabling control over the laser frequency and output power at megahertz frequencies. Outperforming alternative approaches in terms of modulation speed and compactness, this platform holds strong potential for high-speed optical communication applications.

\tocless\section{Results}\label{sec_results}
\subsection*{Tunable external cavity concept}\label{concept}
\begin{figure}[h!]
    \centering
    \includegraphics[width=1.0\linewidth]{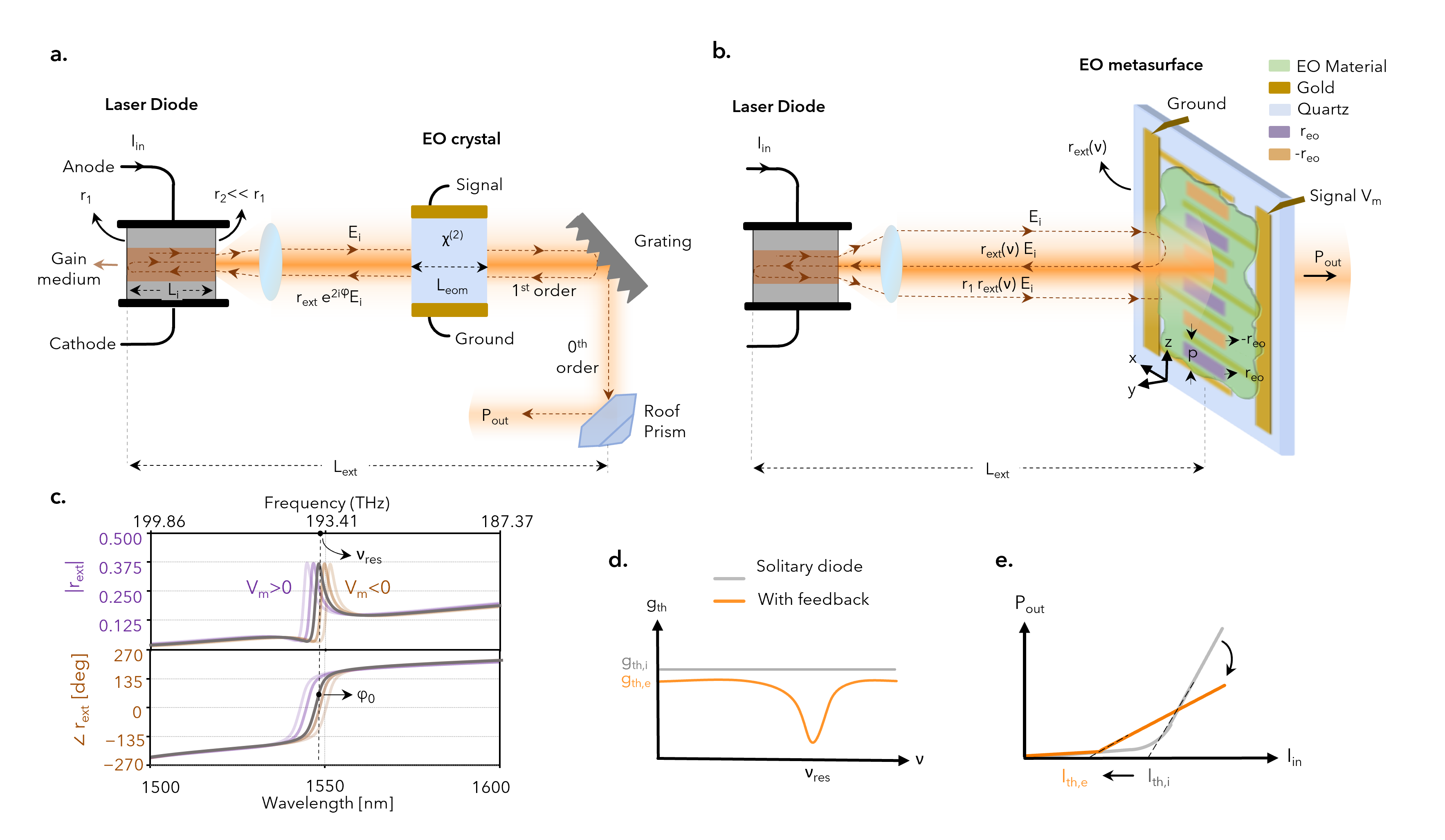}
    \caption{\small
    \setlength{\baselineskip}{0.7\baselineskip}%
    \textbf{Working principle of the metasurface assisted tunable external cavity.} \textbf{a}. The injection current $I_{\mathrm{in}}$ supplies carriers to the gain medium of the laser diode. The back facet has reflection $r_1$, while the front facet is anti-reflection coated with $r_2 \ll r_1$. An EO crystal is placed in the external cavity and modulates the optical path length via the Pockels effect. An applied electric field to the EO crystal induces a refractive index change in the crystal, which results in a laser frequency shift. The optical field $E_i$ inside the cavity is partially transmitted, modulated twice by the EO material, and diffracted by the grating as $r_{\mathrm{ext}}e^{2i\varphi}E_i$, where $\varphi$ is the EO phase shift induced by the modulator. The first-order diffraction provides coherent feedback to the cavity, while the zeroth-order serves as the laser output. The frequency shift scales inversely with the total cavity length $L_{\mathrm{ext}}$, and directly with the EO crystal length $L_{\mathrm{eom}}$ and material parameters of the crystal, such as refractive index and electro-optic coefficient. \textbf{b}. Schematic of the metasurface-integrated external-cavity diode laser. The emitted light is collimated and reflected back by the electro-optic metasurface, establishing external optical feedback. The reflection from the metasurface is frequency-selective, exhibiting a resonance around $\nu_{\mathrm{res}}$, where it reaches a maximum value $r_{\mathrm{max}} \gg r_1$. The reflected beam, $r_{\mathrm{ext}}(\nu) E_i$, re-enters the laser cavity and is reflected by the internal mirror ($r_1$), completing the round-trip feedback path. The metasurface consists of interdigitated gold electrodes and a nonlinear optical polymer, supporting guided-mode resonances. The electro-optic response is activated by periodic poling, producing alternating domains with electro-optic coefficients $+r_{\mathrm{eo}}$ (purple) and $-r_{\mathrm{eo}}$ (orange), aligned through molecular dipole orientation. This configuration enables dynamic modulation of the metasurface reflection via an applied RF field $V_m$. \textbf{c}. The simulation illustrates the resonant behavior of the electro-optic metasurface. The upper graph shows the reflection amplitude, while the lower graph displays the corresponding phase response. The maximum reflection ($r_{\mathrm{max}} = 0.375$) occurs near $\nu_{\mathrm{res}} = 193.41~\mathrm{THz}$, corresponding to a wavelength of $\lambda_{\mathrm{res}} = 1549~\mathrm{nm}$. The metasurface geometry used in the simulation features a pitch size $p = 1.04~\mu\mathrm{m}$ and a nonlinear material layer thickness $T_{\mathrm{JRD1}} = 1.7~\mu\mathrm{m}$. Applying a voltage to the metasurface modifies its refractive index  which leads to the shift of the resonance, as illustrated by semi-transparent curves. \textbf{d}. Without optical feedback, the lasing threshold is determined solely by the length and gain bandwidth of the laser diode. When the EO metasurface provides frequency-selective feedback, the effective gain threshold decreases near the resonance, resulting in a lower required injection current for lasing.  \textbf{e}.  A reduced gain threshold lowers the injection current from \( I_{\mathrm{th},i} \) to \( I_{\mathrm{th}} \), but also decreases the output power \( P_{\mathrm{out}} \) due to modified external cavity conditions.
}
    \label{fig1}
\end{figure}

 One of the simplest and most widely used external cavity setups is the Littrow configuration. Here, light emitted from a laser diode is collimated and directed onto a diffraction grating~\cite{ye2007tunable}. The first-order diffracted beam is reflected back into the diode to provide optical feedback, while the zeroth-order beam serves as the output (Fig.~\ref{fig1}a). In this configuration, adjusting the laser's wavelength is possible only by rotating the grating which is slow and requires careful alignment. To overcome these speed limits, earlier ECL designs incorporated bulk EO crystals directly into the external cavity~\cite{ADP,LT002,bLN1994}. In this configuration, applying a voltage across the EO crystal alters its refractive index via the Pockels effect, thereby modifying the optical path length of the cavity and consequently the lasing frequency. The corresponding frequency tuning can be expressed as $\displaystyle\Delta\nu=-\nu\frac{L_{eom}}{L_{ext}}\frac{\Delta n_{eo}}{n}$, where $\displaystyle L_{eom}$ is the length of the EO modulator, $\displaystyle\frac{\Delta n_{eo}}{n}$ the relative refractive index change and $L_{ext}$ is the total cavity length~\cite{yariv1989quantum}. This relation shows that the achievable frequency shift decreases with increasing external cavity length. Consequently, minimizing the size of the external cavity is desirable to maximize the total frequency shift. However, the large size of commercial electro-optic modulators was a roadblock towards achieving short cavity lengths~\cite{ADP}, resulting in applied voltages reaching hundreds of volts to maximize frequency excursion. 

Guided by our goal to achieve large frequency excursion at lower voltages, we reduce the footprint of the external cavity by condensing the bulk EO crystal and the diffraction grating into a single, compact, frequency-selective electro-optic metasurface with custom-tailored reflectivity $r_\textrm{ext} (\nu)$ (Fig.~\ref{fig1}b). In this case, light generated and emitted from the laser diode $E_i$ by applying the injection current $I_{in}$ impinges directly onto the electro-optic device after being collimated by a lens, and is partially reflected back into the cavity $\displaystyle r_\textrm{ext}\left(\nu\right)E_i$. The back-propagating field then couples into the active region of the laser diode, where it is reflected by the internal mirror $\displaystyle r_1 r_\textrm{ext}\left(\nu\right)E_i$, establishing a full round-trip (indicated by the brown dashed line). The reflection from the emitting facet is neglected since its reflectivity $r_2$ is much weaker than the one provided by the electro-optic metasurface.

The electro-optic metasurface consists of a periodic array of gold stripes with pitch size $p$ patterned on a quartz substrate onto which we spin-coat a layer of the nonlinear molecules JRD1~\cite{jrd1ultrahigh} mixed with polymethylmethacrylate (PMMA), similar to our recent demonstration~\cite{beneaelectro} (details about the geometry are given in Supplementary Information Figure \ref{s-fig:tls}a). The molecules exhibit a $\displaystyle\chi^{(2)}$ nonlinearity, replacing thereby the EO-crystal used in the conventional configuration of Fig.~\ref{fig1}a~\cite{LT002,bLN1994,ADP}. The combination of the slab formed by the nonlinear material and the interdigitated gold stripes supports guided-mode resonances~\cite{guidedmodes}, which occur when the diffraction angles of the gold stripes coincide with the total internal reflection angle of the guided modes.  This enables a frequency-selective reflection $\displaystyle r_\textrm{ext}(\nu)$, with a maximum value $r_\textrm{max}$ occurring at a specific resonance frequency $\displaystyle\nu_\textrm{res}$ and corresponding to the resonance wavelength $\displaystyle\lambda_\textrm{res}={c}/{\nu_\mathrm{res}}$, replacing the grating of Fig.~\ref{fig1}a. By engineering the thickness of the slab and the periodicity of the grating, these guided-mode resonances can be so sparse that a single resonance occurs across the entire gain bandwidth of the laser diode, spanning more than 100 nm, with no other resonances nearby (Fig.~\ref{fig1}c). This ensures feedback on a well-defined resonance and prevents mode hopping.  We note that this extremely high free-spectral range (exceeding 10 terahertz) is a unique feature of sub-wavelength resonances, in stark contrast to approaches using high modal number resonators, such as ring resonators, which feature much lower free-spectral ranges (10s to 100s of gigahertz) and consequently need Vernier filtering to prevent mode hopping. Details about the design of the electro-optic modulator can be found in Supplementary Information Sec.~\ref{Supp:simulations-cst-sweeps}.

We utilize the electro-optic effect provided by the JRD1 molecules as a tuning knob for achieving voltage-controlled laser emission. However, upon spin-coating, the nonlinear molecules are not electro-optically active. They are activated through electric field poling, in which the highly polarizable molecules are oriented along a poling field, producing periodically poled domains with electro-optic coefficients alternating between $-r_{eo}$ and $+r_{eo}$ in adjacent periods. This requires applying a strong DC voltage to the molecules via the interdigitated arrays and heating the sample above glass transition temperature (the full protocol is provided in the Supplementary Information Sec.~\ref{supp:subsec:poling}). After activation, applying an RF field of voltage $\displaystyle V_m$ to the interdigitated electrodes tunes the refractive index $\displaystyle n_{nlm}$ of the entire external mirror by $\displaystyle\Delta n_{eo}(t)=\frac{1}{2}n_{nlm}^3r_{eo}\frac{V_m (t)}{p}$~\cite{beneaelectro} through the Pockels effect. This, in turn, allows controlling the phase and magnitude of the reflection for a given frequency $\displaystyle\nu$ (Fig.~\ref{fig1}c and Supplementary Information Note~\ref{Supp:sec:Modeling of laser behavior} and Fig.~\ref{s-fig:dn}).

 In line with laser theory, the magnitude and the frequency dependence (Fig.~\ref{fig1}c) of the reflection of the electro-optic device, $r_{ext}(\nu)$, influence several key lasing parameters. First, only those modes that experience sufficiently high reflection start lasing, showcasing how a resonant design can achieve narrowband lasing. This is because from the viewpoint of the laser diode, a narrowband reflection lowers the effective gain threshold only in a custom tailored band, in contrast to the solitary anti-reflection-coated laser diode which has a high gain threshold across the entire gain bandwidth (orange and gray curves in Fig.~\ref{fig1}d). Second, the magnitude of the reflection peak determines how much the threshold current, $\displaystyle I_{\text{th}}$, decreases compared to the solitary laser diode since $ \displaystyle I_{\text{th}}(\nu)  \propto \frac{-\ln(r_1 r_{ext}(\nu))}{2L_\mathrm{i}} $, where $L_\textrm{i}$ is the length of the internal cavity, $\displaystyle r_1$ is the reflectivity of the laser facet. The threshold current describes the minimal current at which lasing occurs. Third, the magnitude of the reflection peak also determines the amount by which the slope efficiency $\displaystyle\eta_{\text{slope}}$ decreases (Fig.~\ref{fig1}e),  since $ \displaystyle\eta_{\text{slope}, I>>I_{th}} \propto (1 - r_1 r_\textrm{ext}(\nu)) $. The slope efficiency describes the increase of output power versus current~\cite{milonni_eberly_lasers_ch3to6}. Consequently, while the feedback improves the lasing conditions, it also reduces the output power per unit of injection current above the threshold.

\subsection*{Spectral shaping}\label{Spectral shaping}

\begin{figure} [h!]
\centering
\includegraphics[width=1\linewidth]{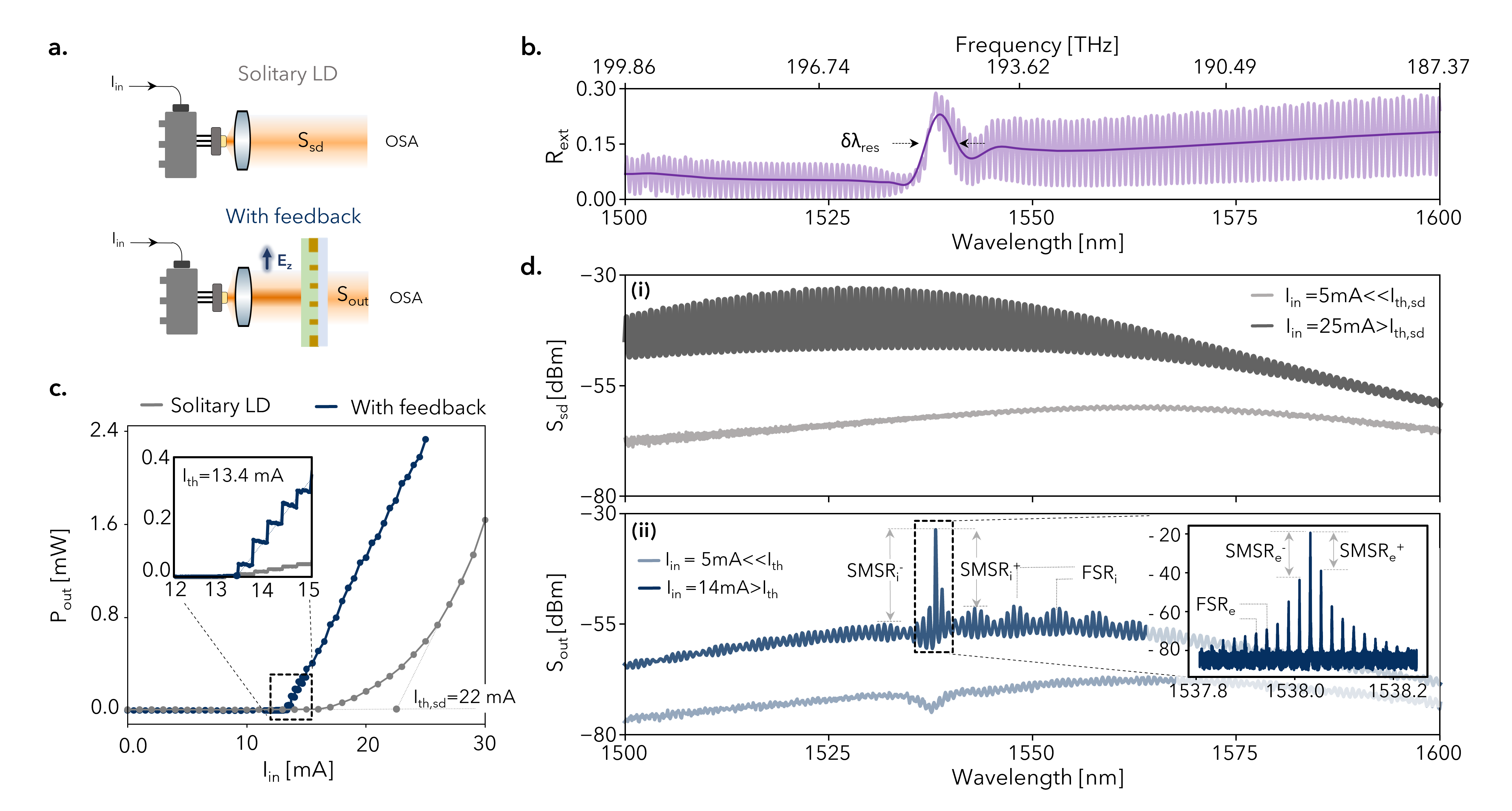}
\caption{\small
\setlength{\baselineskip}{0.7\baselineskip}%
\textbf{Metasurface-controlled laser diode feedback.}
\textbf{a}. To characterize the solitary laser diode, the optical spectrum is sent directly to an Optical Spectrum Analyzer (OSA), and the output power is monitored with a power meter.  The electro-optic metasurface is then placed in the external cavity and aligned such that the electric field inside the cavity is oriented perpendicular to the gold stripes of the metasurface. \textbf{b.} The reflectance spectrum of the metasurface is plotted. The maximum reflectance, $R_{\mathrm{max}} = 0.28$, occurs at the resonance wavelength, with a linewidth of $\delta\lambda_{\mathrm{res}} = 7~\mathrm{nm}$. Fabry–Pérot oscillations from the substrate are visible in light purple, while the filtered reflectance—removing these oscillations—is shown in darker purple for clarity. \textbf{c.} The resulting LI (light–current) curves are compared for the two configurations. The solitary diode shows a lasing threshold of $I_{\mathrm{th,sd}} = 22~\mathrm{mA}$ (gray curve). Once optical feedback is established, the lasing threshold is reduced to $I_{\mathrm{th}} = 13.4~\mathrm{mA}$. The step-like behavior observed in both LI curves originates from mode hopping, which is characteristic of multi-mode laser operation. \textbf{d}. Spectral response with and without feedback.
(i) The emission spectrum of the solitary laser diode, denoted $S_{\mathrm{sd}}$, is recorded using an Optical Spectrum Analyzer. Below threshold (light gray), at $I_{\mathrm{in}} = 5~\mathrm{mA}$, and above threshold, at $I_{\mathrm{in}} = 25~\mathrm{mA}$, the laser exhibits spontaneous and stimulated emission, respectively. The gain spectrum is approximately $80~\mathrm{nm}$ wide, with a noticeable blue shift of $\sim 32~\mathrm{nm}$ as the current increases. (ii) The emission spectra in the presence of feedback, denoted $S_{\mathrm{out}}$, are shown both below threshold at $I_{\mathrm{in}} = 5~\mathrm{mA}$ (light blue) and above threshold at $I_{\mathrm{in}} = 14~\mathrm{mA}$ (dark blue). Below threshold, the resonance of the metasurface is observable in transmission. Above threshold, lasing occurs predominantly around the resonance, while off-resonant modes are suppressed. The internal cavity mode spacing is marked as $\mathrm{FSR}_i$, and the difference in intensity between the main lasing peak and the adjacent internal sidebands is quantified as $\mathrm{SMSR}_i^+$ and $\mathrm{SMSR}_i^-$. A zoomed-in view of the dominant lasing peak reveals the external cavity mode structure, with a free spectral range of approximately $\mathrm{FSR}_e = 5~\mathrm{GHz}$. The relative strength of these external sidebands is characterized by $\mathrm{SMSR}_e^+$ and $\mathrm{SMSR}_e^-$. Supplementary Information Note \ref{Supp:sub:sbr} presents the evolution of the internal sideband ratio $\mathrm{SMSR}_i = \min(\mathrm{SMSR}_i^+, \mathrm{SMSR}_i^-)$ and the external sideband ratio $\mathrm{SMSR}_e = \min(\mathrm{SMSR}_e^+, \mathrm{SMSR}_e^-)$ as a function of injection current. Notably, $\mathrm{SMSR}_i$ remains stable and below $-40~\mathrm{dB}$ above threshold, while $\mathrm{SMSR}_e$ remains below $-20~\mathrm{dB}$ across a wide current range, until disrupted by mode hopping due to the continuous blue shift.}
    \label{fig2}
\end{figure}

Guided by these considerations, we now investigate the impact of the design of the electro-optic device on the electrical and optical characteristics of the external cavity laser by comparing it to the solitary laser diode (Fig.~\ref{fig2}a). We select a device with a pitch size of $p = 1.04~\mu$m, which provides a reflectance whose peak aligns well with the gain peak of the laser diode. The peak reflectance is $R_{\max} \approx 0.28$ at $\lambda_{\mathrm{res}} = 1538$~nm and the linewidth of the resonance is $\delta\lambda_{\mathrm{res}} \approx 8$~nm (Fig.~\ref{fig2}b). Using this device in the external cavity configuration lowers the threshold current significantly to $I_{\text{th}} = 13.4~\mathrm{mA}$, compared to the solitary laser diode with $I_{\text{th,sd}} = 22~\mathrm{mA}$, indicating successful optical feedback between the laser diode and the metasurface (Fig.~\ref{fig2}c). Furthermore, the threshold is markedly more abrupt in the external cavity configuration, as expected. Step-like features appear in the light-current curve in both configurations due to mode hopping: as the injection current varies, the longitudinal modes of the laser shift until a different mode satisfies the resonance condition, producing small yet abrupt changes in output power, in line with existing literature~\cite{gray1991bistability,ogasawara1986output,milonni_eberly_lasers_ch3to6,rauch2015compact,jimenez2017narrow} (inset of Fig.~\ref{fig2}c). In the external cavity configuration, these hops occur when the internal and external cavity modes detune and then re-align. 

The spectral characteristics of laser emission are also markedly different for the two configurations (Fig.~\ref{fig2}d). First, the AR-coated solitary laser diode features multimode emission (Fig.~\ref{fig2}d-i), while the external cavity laser features single-mode operation (Fig.~\ref{fig2}d-ii). Second, while the most intense modes follow the gain peak in the case of the solitary laser diode, this is not the case for the external cavity configuration where the lasing frequency aligns with the maximum of the reflection provided by the electro-optic device, as visible when comparing the reflectance peak in Fig.~\ref{fig2}b and the lasing mode in Fig.~\ref{fig2}d-ii. We note that below the threshold, the resonance of the electro-optic device appears as a dip in the luminescence spectra, since less power is out-coupled from the cavity in this band (gray curve in Fig.~\ref{fig2}d-ii). Third, this frequency-locking is observed over a wide range of currents (see Supplementary Information Note \ref{Supp:sub:sbr}), and mitigates the typical blue-shift of the lasing peak, which is present in laser diodes featuring broadband reflection characteristics (see blue-shift between peak of luminescence spectra below threshold and peak of lasing spectra above threshold for the solitary laser diode in Fig.~\ref{fig2}d-i and Fig.~\ref{s-fig:iv_spectra}~b).  This well-known behavior in III–V semiconductor lasers is driven by the combined effects of band-filling and quasi-Fermi level separation dynamics~\cite{bluegan,chow1999semiconductor}. Altogether, this showcases that our approach combines the advantages of conventional external-cavity diode configurations, such as Littrow and Littman~\cite{ye2007tunable} in frequency locking, while offering a compact solution.

The linewidth of the resonance determines, together with the free spectral range of the external cavity, the degree of suppression of side modes, known as side mode suppression ratio (SMSR). For an effective suppression, the linewidth needs to be commensurate with the free spectral range. For a mode to exist in the external cavity, the total phase accumulation $\varphi_{tot}$ during one round trip in the cavity is an integer multiple of $2\pi$, as
\begin{equation}
    \varphi_\mathrm{tot}=2k_0(n_cL_i+L_\mathrm{ext})+\varphi_0=2\pi m
    \label{eq:FP}
\end{equation}
leading to mode formation at frequencies $\displaystyle\nu_\mathrm{m} = \frac{(m-\frac{\varphi_0}{2\pi}) c_0}{2(n_c L_i + L_\mathrm{ext})}$, where $n_c$ is the refractive index of the cavity, $k_0$ is the vacuum wavenumber, $\varphi_0$ is the phase at the resonance peak of the metasurface and $m$ is the mode number. We choose an external cavity length of $L_{ext} = 2.5~$cm, corresponding to external cavity modes that are mutually separated by approximately $FSR_e = 5~\mathrm{GHz}$. This cavity length ensures that the laser operates in single mode with a minimal SMSR of external modes of $\mathrm{SMSR_e} \approx 25~\mathrm{dB}$, as revealed in the zoom-in of Fig.~\ref{fig2}d-ii. The internal cavity modes with a free spectral range of $FSR_i = 90~\mathrm{GHz}$ are suppressed by up to $\mathrm{SMSR_i} \approx 40~\mathrm{dB}$ (inset of Fig.~\ref{fig2}d-ii). In Supplementary Information Note \ref{Supp:sub:sbr}, we show that these internal and external sideband suppression ratios, $\mathrm{SMSR}_i$ and $\mathrm{SMSR}_e$, are consistent across various injection currents, ensuring controlled laser operation across a large dynamic range.

The exact lasing wavelength can be controlled by adjusting the pitch size of the interdigitated electrode array, with smaller pitch sizes leading to a blue shift in the laser emission (see Supplementary Information Note \ref{supp:sub:CST_psweep}). We showcase this tunability by fabricating and measuring two different electro-optic devices with pitch values of $p = 1.04~\mathrm{\mu m}$ and $p = 1.06~\mathrm{\mu m}$, featuring $\lambda_{\mathrm{res}} = 1539~\mathrm{nm}$ and $\lambda_{\mathrm{res}} = 1559~\mathrm{nm}$, respectively (see Supplementary Information Note \ref{Supp:sub:s shaping}). We chose these values to study the impact of detuning of the electro-optic device's resonant frequency with the gain maximum on the optical and electrical characteristics of the laser. When comparing these two devices, we find a greater reduction in threshold current for the devices with $\lambda_{\mathrm{res}} = 1559~\mathrm{nm}$, although its reflectance is lower. This demonstrates that feedback effectiveness depends not only on the reflectance magnitude, but also on the spectral alignment between the external cavity resonance and the laser diode gain profile. 

We model this behavior by solving the rate equations for a wavelength-dependent reflectivity (Supplementary Information Note \ref{supp:sub:modeling_lres-sweep}).  By studying various detuning conditions between the cavity resonant frequency $\nu_{\mathrm{res}}$ and the central gain frequency $\nu_0$, we confirm that the threshold current is lowest and the output power is highest at zero detuning. 

Overall, our observations demonstrate the importance of co-designing the electro-optic device, the laser cavity and the laser diode for achieving optimal electrical and optical characteristics of the entire system.

\subsection*{Electro-optic frequency tuning}
\label{Electro-optic frequency tuning} 

Building on the static spectral control provided by the metasurface, we now exploit its dynamic tuning capability via the electro-optic effect.
Upon applying an RF voltage $V_m$ to the metasurface, its refractive index is modified by $\Delta n_{\mathrm{eo}}$, resulting in a shift of its resonance frequency $\Delta\nu_{\mathrm{res}}$ of
\begin{equation}
    \frac{\Delta\nu_{\mathrm{res}}}{\nu_{\mathrm{res}}}
    = -\frac{\Delta\lambda_{\mathrm{res}}}{\lambda_{\mathrm{res}}}
    = -\frac{\Delta n_{\mathrm{eo}}}{n_{\mathrm{nlm}}} \Gamma_c
    \label{eq:EOshift}
\end{equation}
where $\Gamma_c$ is the overlap factor between the optical field and the RF field within the nonlinear material~\cite{beneaelectro}. In the small signal approximation, this shift in the metasurface resonance influences laser emission through modifying the reflection phase at the external mirror $\varphi_0$, and hence the round-trip condition at a given wavelength, rather than the absolute value of the reflection $r_{max}$, and hence the gain threshold. This is because the operation point of the laser coincides with the point of maximal reflection, where the roundtrip losses are lowest. Here, the rate of change of the phase with voltage $\displaystyle\frac{\partial \varphi_0}{\partial V_m} = \frac{\partial \varphi}{\partial V_m}|_{\nu_{res}}= \frac{\partial \varphi}{\partial \nu_{res}}\frac{\partial \nu_{res}}{\partial V_m}|_{\nu_{res}}$ is maximal since $\frac{\partial \varphi}{\partial \nu_{res}}|_{\nu_{res}}$ is maximal while the rate of change of the magnitude $\displaystyle\frac{\partial r_{ext}}{\partial V_m}|_{\nu_{res}} = \frac{\partial r_{ext}}{\partial \nu_{res}}\frac{\partial \nu_{res}}{\partial V_m}|_{\nu_{res}}$ is zero since $\frac{\partial r_{ext}}{\partial \nu_{res}}|_{\nu_{res}} = 0$. This fundamental property of the metasurface controlled external cavity laser we propose is essential, as it results in a frequency modulation of laser emission, and minimizes intensity modulation. Such decoupling of frequency modulation from intensity modulation is highly desirable in most applications, including communications and ranging applications. The exact amount of amplitude modulation is measured in the last section of this manuscript. 

\begin{figure}[t!]
    \centering
    \includegraphics[width=1.0\linewidth]{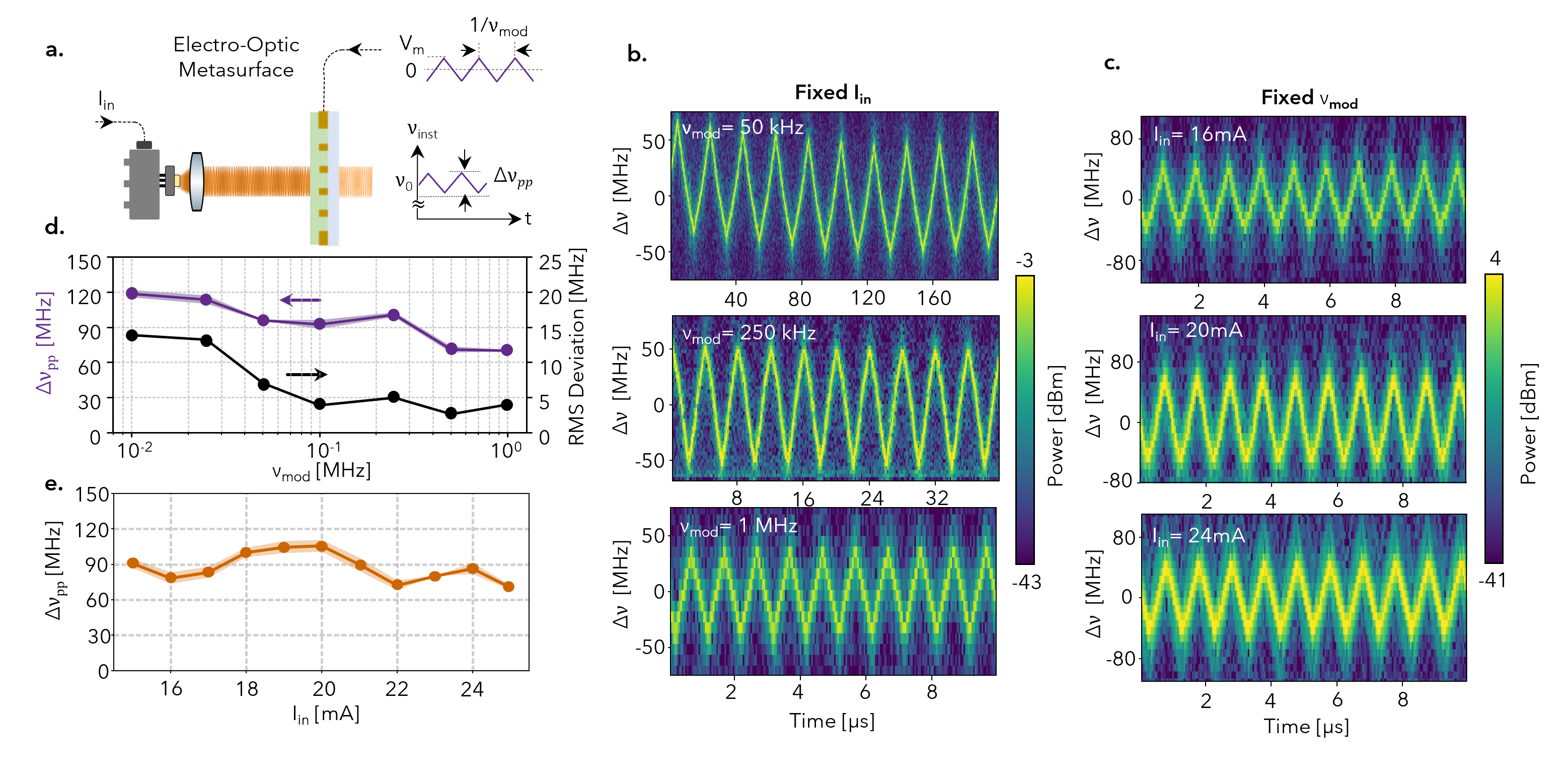}
    \caption{\small
    \setlength{\baselineskip}{0.7\baselineskip}%
    \textbf{Ultrafast frequency modulation of a laser diode with an electro-optic metasurface}. \textbf{a}. An RF source applies a triangular voltage waveform with peak amplitude $V_m$ and frequency $\nu_{\mathrm{mod}}$ to the metasurface, dynamically tuning its reflection. This modulation alters the lasing frequency of the diode laser. The output beam is combined with a frequency-stable reference laser in a heterodyne detection setup, enabling extraction of the instantaneous frequency $\nu_{\mathrm{inst}}(t)$ through RF beatnote analysis. The modulation induces a frequency excursion $\Delta \nu$ centered around a carrier frequency $\nu_0$. The detailed setup is presented in Supplementary Information Figure ~\ref{s-fig:HD_eo_setup}.
    \textbf{b}. Spectrograms of the heterodyne beatnote signal under triangular RF modulation at three different frequencies: $50~\mathrm{kHz}$, $250~\mathrm{kHz}$, and $1~\mathrm{MHz}$. All measurements were performed at a fixed RF amplitude of $V_m = 10~\mathrm{V}$ and injection current $I_{\mathrm{in}} = 16~\mathrm{mA}$. Each panel shows ten consecutive modulation periods. An average frequency excursion of approximately $100~\mathrm{MHz}$ is observed across all modulation rates. The color scale indicates optical power, calibrated to the output power from the laser’s LI curve. The spectrograms capture the time-resolved evolution of the laser frequency enabled by the electro-optic metasurface. Even at the highest modulation rate, the excursion remains visible, demonstrating the fast response of the metasurface-enabled tuning. At higher modulation frequencies, however, the intrinsic time–frequency resolution trade-off of spectrogram analysis limits the ability to fully resolve the instantaneous frequency.
    \textbf{c}. Heterodyne spectra measured at a fixed modulation frequency $\nu_{\mathrm{mod}} = 1~\mathrm{MHz}$ for injection currents $I_{\mathrm{in}} = 16~\mathrm{mA}$, $20~\mathrm{mA}$, and $24~\mathrm{mA}$. The traces illustrate how the frequency excursion and heterodyne signal amplitude evolve with injection current. As the current increases, the frequency excursion gradually narrows. The amplitude of the beatnote is calibrated using the laser’s LI curve. \textbf{d.} Frequency excursion versus modulation frequency. The left vertical axis (purple) shows the peak-to-peak frequency excursion $\Delta \nu_{pp}$, with the shaded region representing the standard deviation across modulation cycles. The right axis (black) shows the root-mean-square (RMS) deviation from an ideal triangular waveform. As $\nu_{\mathrm{mod}}$ increases, both the frequency excursion and the waveform fidelity degrade, illustrating the trade-off between modulation speed and dynamic range. \textbf{e.} Frequency excursion as a function of injection current $I_{\mathrm{in}}$ at fixed modulation frequency $\nu_{\mathrm{mod}} = 1~\mathrm{MHz}$. The shaded area indicates the standard deviation of the measured excursion. The frequency excursion remains relatively stable around $90~\mathrm{MHz}$ for moderate injection currents, with a slight decrease observed at higher current levels.
    }
    \label{fig3}
\end{figure}
The amount of frequency modulation $\Delta \nu_\mathrm{m,V=V_m}$ at voltage $V_m$ can be retrieved by re-evaluating the round-trip condition upon a change in the phase of the reflected electric field (from its original value $\varphi_0$ to $\varphi_\mathrm{0} +\Delta\varphi_\mathrm{eo} $):
\begin{equation}
    \Delta \nu_\mathrm{m,V=V_m}=\frac{c_0\Delta\varphi_\mathrm{eo}}{4\pi(n_{c}L_i+L_\mathrm{ext})}
    \label{eq:lasing_shift}
\end{equation}

Using the relationships above, the amount of phase modulation $\displaystyle\Delta \varphi_{eo}$ depends on the shift of the resonance $\Delta \nu_{res}$ and on the steepness of the phase modulation around the resonance $\displaystyle\frac{d\varphi}{d\nu_{res}}|_{\nu_{res}}$, yielding $\displaystyle\Delta\varphi_\mathrm{eo}\approx\frac{d\varphi}{d\nu_{res}}|_{\nu_{res}}\Delta\nu_\mathrm{res}$. This underlines the importance of having both high quality factor resonances (yielding large steepness) and a strong electro-optic response (yielding large shifts of the resonance frequency upon an applied bias) for achieving large frequency excursions. We quantify these aspects using finite element methods (see Supplementary Note~\ref{supp:sub:phase modulation-sweep}).
Together, these considerations result in an EO-induced shift in the lasing frequency that depends linearly on the shift in the resonance frequency, $\displaystyle\Delta\nu_\mathrm{res}$, and hence in applied voltage $V_m$, and inversely on the length of the external cavity, $L_\mathrm{ext}$ (see Methods):
\begin{equation}
    \Delta \nu_\mathrm{m,V=V_m}
    =
    \frac{c_0}{4\pi(n_{c}L_i+L_\mathrm{ext})}\frac{d\varphi}{d\nu_{res}}|_{\nu_{res}}g_\mathrm{eo}V_m,
    \label{eq:lasinglinear}
\end{equation}

where $\displaystyle g_\mathrm{eo}=\frac{\nu_{\mathrm{res}}\, n_{\mathrm{nlm}}^{3} r_{\mathrm{eo}} \Gamma_c}{2d}$, is a proportionality constant that characterizes the electro-optic response of the nonlinear material to the applied $V_m(t)$. A higher quality-factor resonance is beneficial, as its narrower optical linewidth $\delta\nu_{\mathrm{res}}$ supports larger frequency excursion at lower values of the RF modulation voltage $V_m$. 

We demonstrate electro-optic resonance tuning by driving the metasurface with a triangular RF voltage of amplitude $V_m = 10~V$ and frequency $\nu_{mod}$ in the configuration shown in Fig.~\ref{fig3}a. The laser’s instantaneous frequency response is monitored via heterodyne detection. The time-resolved frequency excursion $\Delta \nu(t)$ is extracted from the beating between the modulated laser and a reference laser. We find that the instantaneous laser frequency follows closely the triangular shape of the modulation voltage, as illustrated in the spectrograms obtained for modulation frequencies of $\nu_{mod} = 50~\mathrm{kHz}$, $250~\mathrm{kHz}$, and $1~\mathrm{MHz}$ (Fig.~\ref{fig3}b). This linear dependence on applied voltage is consistent with Eq.~\ref{eq:lasinglinear}. For an applied peak voltage amplitude of $V_m = 10~\mathrm{V}$ and constant injection current $I_{\mathrm{in}} = 16~\mathrm{mA}$, frequency excursions of approximately $80~\mathrm{MHz}$, $90~\mathrm{MHz}$, and $60~\mathrm{MHz}$ are observed at these three modulation frequencies, respectively. These values remain consistent across various injection currents, including further above threshold: $I_{\mathrm{in}} = 16~\mathrm{mA}$, $20~\mathrm{mA}$, and $24~\mathrm{mA}$, for a modulation frequency of $\nu_{\mathrm{mod}} = 1~\mathrm{MHz}$, where the frequency excursion remains nearly constant between $70$ and $90~\mathrm{MHz}$ (Fig.~\ref{fig3}c). The absolute optical powers of the laser emission are retrieved by calibrating the intensity of the heterodyne signal with the IV-curve measurement shown in Fig.~\ref{fig2}c.

When analyzing the peak-peak frequency excursion $\Delta \nu_{pp}$ as a function of modulation frequency, somewhat larger excursions are observed at lower modulation frequencies (purple curve, Fig.~\ref{fig3}d). Furthermore, since an accurate prediction of the instantaneous frequency is highly relevant in practical applications, we investigate the root-mean-square (RMS) deviation of the instantaneous frequency from the ideal triangular waveform (see right axis of Fig.~\ref{fig3}d and Supplementary Information Note ~\ref{Supp:sub:rms}). While at higher modulation frequencies, this deviation is below 5 MHz, it increases at lower modulation frequencies due to instabilities known to originate from the mechanics of the external cavity. We note that the reduced spectral resolution decreases the accuracy of the RMS calculation at higher frequencies. At a constant modulation frequency of $\nu_{mod}= 1~\mathrm{MHz}$, the excursion remains between $70$ and $120~\mathrm{MHz}$ for various injection currents, as shown in Fig.~\ref{fig3}e. Shaded regions indicate the standard deviation. These results are in line with estimated values of the frequency excursion for our particular metasurface design, demonstrating the validity of our approach (see Supplementary Information Note \ref{supp:sub:phase modulation-sweep}). 

\subsection*{Electro-optic intensity modulation at the threshold}\label{Electro-optic intensity modulation at the threshold} 
Beyond frequency control, electro-optic driving of the metasurface also enables modulating the laser intensity when operated near the threshold. This effect arises because dynamically shifting the resonance creates a time-dependent detuning between the laser gain and the feedback provided by the metasurface at a fixed optical frequency. Since at threshold, gain is just sufficient to balance for losses for a single mode $\displaystyle g_{th} = -\frac{ln(r_1r_{ext}(\nu_{res}))}{2L_i}$, so the transient detuning can temporarily alter this balance - effectively turning on and off the laser diode and introducing intensity modulation. Above threshold, where the gain significantly overcomes losses, this intensity modulation is expected to be significantly suppressed, in line with our discussion above. 

\begin{figure}[h!]
    \centering
    \includegraphics[width=0.85\linewidth]{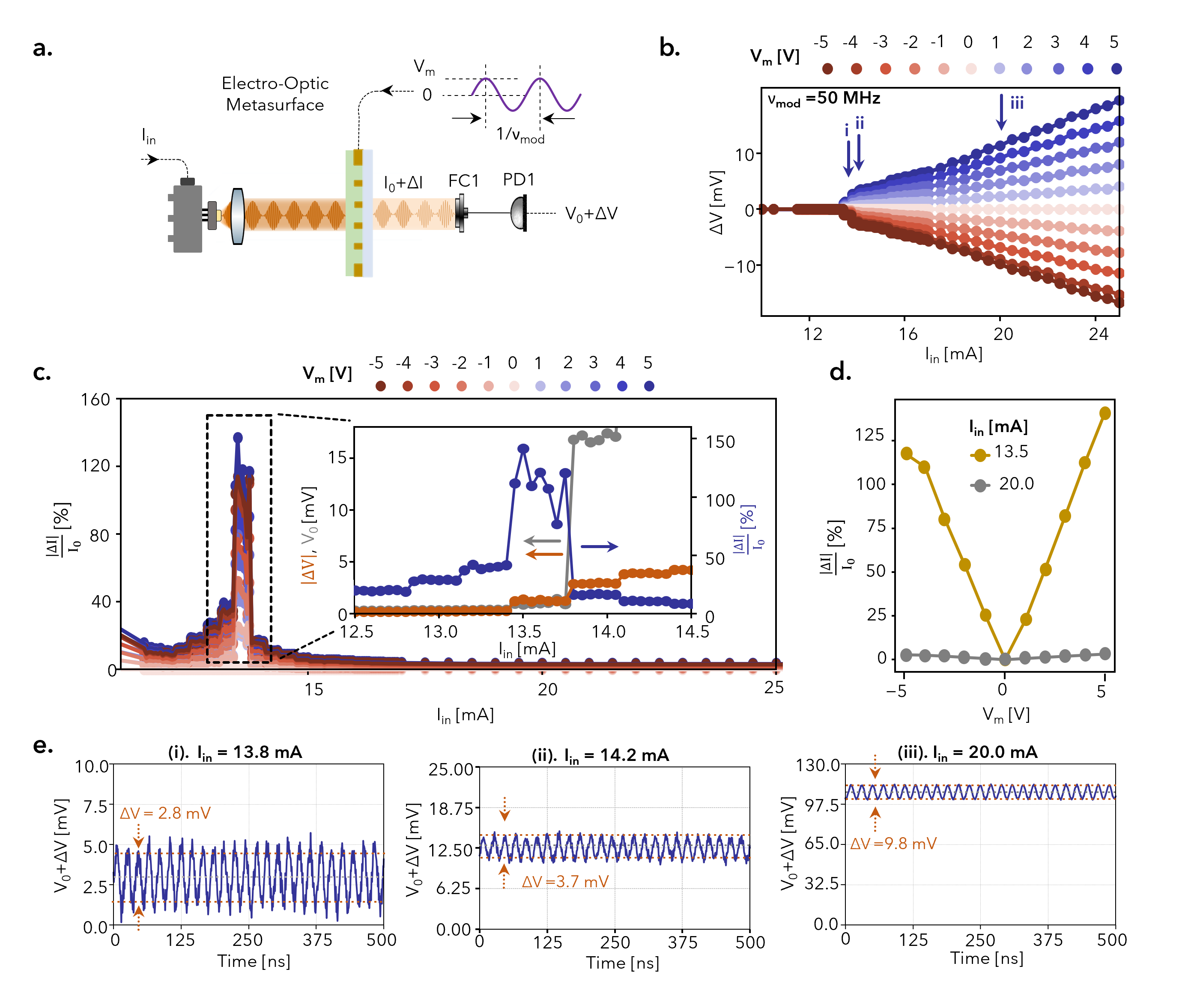}
    \caption{\small
    \setlength{\baselineskip}{0.7\baselineskip}%
    \textbf{Experimental setup for amplitude modulation.} \textbf{a.} After optical feedback is established between the laser diode and the electro-optic metasurface, an RF voltage with amplitude $V_m$ at $\nu_{mod} = 50~\mathrm{MHz}$ is applied using an arbitrary waveform generator. This signal is also sent as a reference to the lock-in amplifier. The detailed setup is presented in Supplementary Information Figure ~\ref{s-fig:HD_eo_setup}. After coupling the output light from free space into a fiber using fiber coupler FC1 and directing it to photodiode PD1, the detected signal appears as $V_0 + \Delta V$, consisting of a DC component $V_0$ and an RF component $\Delta V$ oscillating at $\nu_{mod}$. In the homodyne detection path, the lock-in amplifier measures the modulated signal $\Delta V$, while the DC component $V_0$ is recorded from the DC output of PD1 using a voltmeter.  \textbf{b}. Lock-in measurements of the modulation. The plots show the variation of the detected voltage $\Delta V$ as a function of injection current $I_{\mathrm{in}}$ and RF amplitude $V_m$, measured at a modulation frequency of $\nu_{mod} = 50~\mathrm{MHz}$. The bipolar colorbar indicates the magnitude and sign of $\Delta V$. \textbf{c}. Measured modulation efficiency $\Delta I / I_0$ as a function of injection current $I_{\mathrm{in}}$ for various RF amplitudes $V_m$ at $\nu_{mod} = 50~\mathrm{MHz}$. The efficiency peaks near the laser threshold, exceeding $100\%$ for $V_m > 4~\mathrm{V}$. Inset: Zoomed-in view showing the absolute voltage variation $\Delta V$ (orange), the DC component $V_0$ (gray), and the corresponding modulation efficiency $\Delta I / I_0$ (blue, right axis) at $V_m = 5~\mathrm{V}$. Mode-hopping events are also visible as discrete jumps in the traces. For background on the threshold behavior, see Supplementary Information Figure \ref{s-fig:sbr}. \textbf{d}. Maximum modulation efficiency plotted for two regimes: injection currents near threshold $I_1 = 5~\mathrm{mA}$ (gold) and well above threshold $I_3 = 20~\mathrm{mA}$ (gray). The results show that the maximum modulation increases approximately linearly with the RF amplitude $V_m$ in both regimes. \textbf{e}. Oscilloscope time-trace measurements at fixed voltage ($V_m = 5$~V) and varying currents, as indicated by the arrows in panel b. The vertical axis shows the RF modulation superimposed on the DC signal $V_0$. Traces are shown near threshold at $I_\mathrm{in} = 13.8~\mathrm{mA}$, slightly above threshold at $I_\mathrm{in} = 14.2~\mathrm{mA}$, and well above threshold at $I_\mathrm{in} = 20~\mathrm{mA}$. As the output power increases, the modulation depth decreases due to the faster growth of the DC component $V_0$ compared to the modulated signal $\Delta V$.
    }
    \label{fig4}
\end{figure}

To demonstrate amplitude modulation, we operate the external cavity system discussed above close to its threshold and monitor the output power using the setup schematically illustrated in Fig.~\ref{fig4}a. A detailed view of the experimental configuration is provided in Supplementary Information Fig.~\ref{s-fig:HD_eo_setup}. A sinusoidal RF signal with amplitude $V_m$ and frequency $\nu_{mod} = 50~\mathrm{MHz}$ is applied to the metasurface, while the laser output intensity $I_0+\Delta I$ is detected by a photodiode yielding an electrical signal $V_0 + \Delta V$, where $V_0$ is the DC component and $\Delta V$ is the RF component oscillating at the modulation frequency $\nu_{mod}$. When sweeping the current from below to above threshold, we find that the absolute intensity modulation is increasing, exhibiting a symmetric response for positive and negative drive voltages (Fig.~\ref{fig4}b). However, in many applications, the relative intensity modulation is of highest relevance. Consequently, we quantify the modulation efficiency $\Delta I / I_0 =\Delta V / V_0 $ for various injection currents $I_{\mathrm{in}}$ and RF amplitudes $V_m$ and find that the efficiency peaks sharply near the lasing threshold and exceeds $100\%$ for drive amplitudes higher than $4\mathrm{V}$ (Fig.~\ref{fig4}d). This highlights the enhanced susceptibility of the intracavity dynamics to reflection modulation when the laser is operating close to threshold. The inset provides a zoomed-in view of the region near threshold for $V_m = 5~\mathrm{V}$, where the absolute variation $\Delta V$ (orange), the DC voltage $V_0$ (gray), and their ratio $ \Delta I / I_0$ (blue, right axis) are shown simultaneously. Discrete steps in all traces, indicate mode-hopping behavior as the injection current is varied. When sweeping the RF amplitude from $-5~\mathrm{V}$ to $+5~\mathrm{V}$, we find that the absolute intensity modulation increases approximately linearly with $V_m$ (yellow curve in Fig.~\ref{fig4}d), as expected for the electro-optic effect. Above threshold, the intensity modulation is significantly lower, and remains below 5$\%$ (gray curve in Fig.~\ref{fig4}~d). This is confirmed by directly observing the temporal behavior of the amplitude-modulated laser through recording oscilloscope time traces of the photodetected signal at different injection currents (Fig.~\ref{fig4}~e). The three panels correspond to operation near threshold at $I_\mathrm{in} = 13.8~\mathrm{mA}$, showcasing how the laser is turning on and off (Fig.~\ref{fig4}e-i), slightly above threshold at $I_\mathrm{in} = 14.2~\mathrm{mA}$ (Fig.~\ref{fig4}e-ii), and well above threshold at $I_\mathrm{in} = 20~\mathrm{mA}$. As the injection current increases, the baseline DC component $V_0$ grows significantly, while the amplitude of the modulation $\Delta V$ increases more slowly, leading to a reduced relative modulation depth in the above-threshold regime. Finally, we show that this increased intensity modulation around laser threshold can be reproduced by solving the rate equations (see Supplementary Information Note \ref{supp:sub:modeling_modulation}). Since this approach solves for steady state solutions, it is applicable for the case of slow modulation discussed here, where all laser dynamics occur on significantly faster timescales.

Overall, these results demonstrate that electro-optic driving of the metasurface-feedback laser enables efficient amplitude modulation around threshold, while keeping intensity modulation small above threshold. In this regime, the laser enables efficient frequency tunability. 

\newpage
\tocless\section{Conclusions and outlooks}\label{sec_conclusions}
This work presents the first integration of a resonant electro-optic metasurface into an external cavity diode laser, enabling high-speed control of laser emission through bias voltage. Compared to previous designs requiring multiple bulky components, our platform overcomes the challenges in alignment, footprint and frequency excursion by condensing both the grating and the electro-optic tuning element into a single ultrathin device, significantly reducing the external cavity length. 

By employing high-Q resonances in sub-wavelength thick metasurfaces, we achieve single-frequency feedback across a bandwidth that exceeds 100~nm (>13 THz). This approach stands in stark contrast with alternative approaches that rely on ring resonators as the feedback medium~, which operate on high modal numbers and support modes with typical free-spectral ranges of 100s of gigahertz to 1 THz, making them more vulnerable to mode hopping or requiring Vernier filtering~\cite{guo_thermally_2021}. By carefully designing the properties of the optical resonance such as its resonance frequency and $Q$ factor, we achieve single mode operation across a large current range in a compact design. By choosing an in-plane periodically poled film of highly non-linear molecules JRD1, we achieve efficient modulation of the metasurface refractive index, enabling frequency excursions exceeding 100 MHz and amplitude modulation above 100$\%$ around the threshold. We find that the frequency modulation remains stable across a wide range of injection currents $I_{\mathrm{in}}$, while the amplitude modulation efficiency exhibits a pronounced peak near the lasing threshold. Altogether, these findings establish electro-optic metasurfaces as a powerful platform for active control of laser dynamics in both frequency and amplitude domains. 

Our design guidelines showcase several avenues for custom-tailoring and improving the performance of future metasurface-based external cavity lasers. First, metasurface design relying on quasi-bound states in the continuum~\cite{PhysRevLett.121.193903}, Mie resonances~\cite{lawrence_high_2020} or photonic crystal structures have been shown to achieve narrow linewidths and high reflectivity, opening the possibility to further improve the frequency modulation, laser threshold and side mode suppression ratio. Second, placing the metasurface closer to the output facet of the laser diode would reduce the cavity length and therefore increase the maximal frequency excursion. Owing to the simplicity of fabrication of our proposed devices, they could potentially be embedded directly onto the laser facet. Third, the nonlinear material also offers significant flexibility. On one side, other materials platforms could be envisioned in the future, such as lithium niobate~\cite{klopfer_high-quality-factor_2022, weigand2021enhanced}, hybrid silicon-organic~\cite{soma_subvolt_2025}, III-V materials~\cite{wu2019dynamic} or phase change materials~\cite{zhang2021electrically,sha2024chirality,Barreda:22}. On the other side, the EO material can be poled using electrodes of arbitrary shape, enabling custom-tailoring of their crystallographic properties and, consequently, their electro-optic properties in the two-dimensional plane of the metasurface. This could enable high-speed on-demand  generation of laser modes with non-trivial spatial distribution such as those carrying orbital angular momentum~\cite{jones,modeslectivevecsel}. 

We anticipate that the design guidelines and modeling we provide will help enhance the performance of these metasurface-controlled external cavity lasers. Their compact form factor, simple fabrication, frequency robustness and low power consumption are essential characteristics for establishing electro-optically tunable external cavities as a powerful platform for communication and LiDAR systems~\cite{industrial}. In imaging systems such as microscopy, their high-speed tunability may provide opportunities to perform in-situ spectroscopy by tuning across resonances on time scales faster than molecular dynamics.

\tocless\section{Methods}
\subsection*{Device fabrication and electro-optic activation.} The electro-optic metasurfaces were fabricated on a 1-mm-thick quartz substrate using a standard lift-off process (PMMA-MMA) to define a 50 nm–10 nm gold–titanium (Au–Ti) layer. Subsequently, a solution of the nonlinear material JRD1 mixed with PMMA was prepared for spin-coating onto the substrate. This mixture, consisting of $50~\%wt$ JRD1 powder and PMMA in solid form, was blended with cyclopentanone (10~\%wt) as a solvent for 12 hours using a swirling machine. The resulting solution was filtered through $0.2~\mu\mathrm{m}$ PTFE filters and applied to the chip via a three-step spin-coating process: 500 rpm for 5 seconds, 1500 rpm for 30 seconds, and 1000 rpm for 30 seconds.After spin-coating, the film was hard-baked at $65^\circ$C for 24 hours. A uniform layer thickness of $t_{\mathrm{JRD1}} = 1.8~\mu\mathrm{m}$ was measured after baking. The device was then electro-optically activated through a poling procedure. During poling, an electric field of up to $100~\mathrm{V}/\mu\mathrm{m}$ was applied via the electrodes to the polymeric material under a nitrogen atmosphere, at the poling temperature ($T_p = 84^\circ$C), for 15 minutes to allow molecular alignment with the applied field. The electric field was maintained while the sample cooled down to room temperature.
A comparison of IV curves obtained with a DC source (Keithley 2400) before and after poling shows increased conductivity, as presented in the supplementary Information Figure \ref{s-fig:poling}. Additionally, a change in the film's topography after poling, measured via Atomic Force Microscopy (AFM), confirms the successful alignment of the nonlinear molecules\cite{beneaelectro} (see supplementary Information Figure \ref{s-fig:poling}).

\subsection*{Electro-optic frequency tuning.}
To derive the EO-induced shift in lasing frequency, we start from the resonance condition in Eq.~\ref{eq:FP}, which determines the allowed mode frequencies as 
$\displaystyle\nu_\mathrm{m} = \frac{\left(m - \frac{\varphi_0}{2\pi}\right) c_0}{2(n_c L_i + L_\mathrm{ext})}$.
When an EO voltage $V_m$ is applied, the metasurface reflection phase shifts from $\displaystyle\varphi_0$ to $\displaystyle\varphi_0 + \Delta \varphi_{\mathrm{eo}}$, altering the round-trip phase condition and thus the lasing frequency from $\nu_0$ to $\nu_0+\Delta \nu_m$. Assuming the reflection phase $\varphi$ varies smoothly around the resonance, we linearize its frequency dependence using a first-order Taylor expansion
$\displaystyle\Delta\varphi_\mathrm{eo}\approx\frac{d\varphi}{d\nu_{res}}|_{\nu_{res}}\Delta\nu_\mathrm{res}$
where $\displaystyle\frac{d\varphi}{d\nu_{res}}|_{\nu_{res}}$ is the local slope of the phase spectrum at the lasing frequency (where $r_{ext}$ is maximal) and $\Delta \nu_{\mathrm{res}}$ the shift in the resonance frequency of the metasurface upon applied bias. Inserting this into the round-trip condition gives:
\begin{equation}
\displaystyle\Delta \nu_{\mathrm{m},V=V_m} \approx \frac{c_0}{4\pi(n_c L_i + L_\mathrm{ext})} \frac{d\varphi}{d\nu_{res}}|_{\nu_{res}}\Delta \nu_{\mathrm{res}}.
\label{eq:lasinglinear0}
\end{equation}
The resonance frequency shift is related to the EO-induced index modulation by 
$\displaystyle\Delta \nu_{\mathrm{res}} = \frac{\nu_{\mathrm{res}} \Delta n_{\mathrm{eo}}}{n_{\mathrm{nlm}}} \Gamma_c$, 
yielding
\begin{equation}
\Delta \nu_{\mathrm{m},V=V_m}
= \frac{c_0}{4\pi(n_c L_i + L_\mathrm{ext})} \frac{d\varphi}{d\nu_{res}}|_{\nu_{res}} \frac{\nu_{\mathrm{res}} \Delta n_{\mathrm{eo}}}{n_{\mathrm{nlm}}} \Gamma_c
= \frac{c_0}{4\pi(n_c L_i + L_\mathrm{ext})} \frac{d\varphi}{d\nu_{res}}|_{\nu_{res}} g_{\mathrm{eo}} V_m,
\label{eq:lasinglinear1}
\end{equation}
which corresponds to Eq.~\ref{eq:lasinglinear}.

Starting from this formula, we recognize the ingredients that are necessary to achieve a large frequency excursion $\Delta \nu_{pp}$: a steep slope $\displaystyle\frac{d\varphi}{d\nu_{res}}|_{\nu_{res}}$ at the lasing frequency, a large electro-optic coupling rate $g_{eo}$ and a short cavity, corresponding to small $L_{ext}$. A steeper phase slope $\displaystyle\frac{d\varphi}{d\nu}$ corresponds to an increased group delay $\displaystyle\tau_g = \frac{1}{2\pi} \frac{d\varphi}{d\nu}$, and hence increased photon lifetime $\tau$. Since the group delay is an approximate measure of this photon lifetime in Lorentzian resonators, this is directly related to the quality factor by $\displaystyle Q = 2\pi\nu_{\mathrm{res}} \tau$. More generally, the group delay is bounded by the photon lifetime, i.e., $\displaystyle\tau_g \leq \tau = Q / 2\pi\nu_{\mathrm{res}}$~\cite{grgic2011coupled}, with equality approached only in ideal cases. Consequently, a higher quality-factor resonance translates into a linearly increased frequency excursion for a given RF modulation voltage $\displaystyle V_m$.

\subsection*{Optical setups and measurements} Characterization of the metasurface resonance was performed using the reflection/transmission measurement setup depicted in the Supplementary Information Figure \ref{s-fig:tls}. The wavelength of a continuous-wave (CW) laser (Keysight 8164B) was swept over its full available range, from 1490~nm to 1630~nm, with a resolution around 10~pm. To characterize the resonance of the metasurface, transmission and reflection measurements were conducted using a wavelength-calibrated power meter (Keysight 81624B). The same setup was then used to assess the reflection and transmission modulation strength for both polarization states. An RF voltage with amplitude $\displaystyle V_m = 5~\mathrm{V}$ and modulation frequency $\displaystyle v_m = 50~\mathrm{MHz}$ was generated using an arbitrary waveform generator (Zurich Instruments HDAWG) and sent simultaneously to the device and to a lock-in amplifier (Zurich Instruments UHFLI) for synchronization. The modulation efficiency, expressed as $\displaystyle\eta_R = \frac{\Delta I_R}{I_{\mathrm{max}}}$, was measured for different laser emission wavelengths.

An anti-reflection-coated laser diode (SAL-1550-100) is used as a coherent light source, controlled by the ITC4001 TEC module, which provides the required injection current and maintains a constant temperature of the laser diode. An isolator is placed after the electro-optic metasurface to prevent parasitic reflections from returning to the laser diode. A balanced photodiode (PD1, Thorlabs PDB480C-AC) with a 1.6~GHz bandwidth collects the light for the homodyne detection path. The RF output of the photodiode is connected to the UHFLI lock-in amplifier, while the DC output is connected to a Keithley 2400 Source Meter. All PI curves presented throughout this work were measured using the DC output of PD1 and read via the Keithley meter. The coupling efficiency of FC1, the splitting ratio of the beam splitter, and the transimpedance gain of the photodiode are all taken into account to calculate the absolute power after the device. For spectral shaping measurements~(Fig. \ref{fig2}d-(ii)), an optical spectrum analyzer~(Anritsu MS9740B) was used to measure the emission spectrum. For the zoom-in inset of Fig. \ref{fig2}d-(ii), high-resolution measurements were performed using the Apex AP2043B optical spectrum analyzer. The arbitrary waveform generator was used to modulate the device and to provide the reference signal for the lock-in amplifier. The laser light co-propagates in a 50/50 fiber splitter with a stable CW laser (Keysight N7776C). The output of the splitter is sent to a fast photodiode with 5~GHz bandwidth (PD2, Thorlabs DET08CFC). The CW laser is swept over a range of $\pm 200$~pm around the lasing mode of the laser diode. An Extra604A Keysight oscilloscope records the output of PD2, and the time trace on the oscilloscope is used as the raw data for spectral analysis.

\vspace{1cm}

\textbf{Data Availability}
The data generated within this paper is available together with plotting scripts in the Zenodo database under https://zenodo.org/uploads/17280446.

\medskip
\textbf{Code Availability}
The code used to plot the data within this paper is available in the Zenodo database under https://zenodo.org/uploads/17280446.


\medskip
\textbf{Acknowledgments}
We thank Dr. Alexander Dikopoltsev for fruitful discuassions and Prof. Camille Sophie Brès for providing access to a high-resolution OSA from her lab. Z.B., A.T. and I.C.B.C. acknowledge funding from the Swiss National Science foundation under PRIMA Grant No. 201547. G.J. acknowledges funding from the Swiss National Science Foundation - National Science Foundation (SNSF-NSF) Lead Agency program under award number 219409. The initial sample fabrication, which was critical to establishing the proof-of-concept, was carried out at the Center of MicroNanoTechnology (CMi) at EPFL. We gratefully acknowledge the CMi facilities and the support provided by its staff. Z.B. acknowledges Francesco Bertot for assistance with the first development of the poling procedure. Z.B. acknowledges Dr. Joachim Sacher for insightful discussions about mode hopping of the solitary laser diode.
\medskip

\textbf{Author contributions}
I.C.B.C., Z.B. and A.T. conceptualized the project. A.L. and Z.B. built the experimental setup. Z.B. and A.T. performed the measurements. G.J. developed the heterodyne measurement setup. I.C.B.C and Z.B. fabricated the samples. Z.B performed the CST simulations and analyzed the data. I.C.B.C and Z.B. performed the rate equation simulations and analyzed the data. Z.B.
and I.C.B.C. wrote the manuscript with feedback from all authors. The work was done under the supervision of I.C.B.C.

\medskip
\textbf{Competing interests}
The authors declare no competing interests.

\textbf{Corresponding authors} 
Correspondence Zahra Basiri (zahra.basiri@epfl.ch) and Ileana-Cristina Benea-Chelmus (cristina.benea@epfl.ch). 

\bibliographystyle{paper}
\bibliography{bibliography}

\appendix
\renewcommand{\thefigure}{S\arabic{figure}}
\setcounter{figure}{0} 
\setcounter{linenumber}{1}
\setcounter{equation}{0}
\renewcommand{\theequation}{S\arabic{equation}}
\clearpage 
\vspace*{1cm}

\begin{center}
    {\Huge \bfseries Supplementary Information}
\end{center}

\vspace{1cm}
\tableofcontents
\section{Sample details} \label{supp:sec:Sample details}

\subsection{Poling of the nonlinear material}\label{supp:subsec:poling}
\begin{figure}
    \centering
    \includegraphics[width=0.9\linewidth]{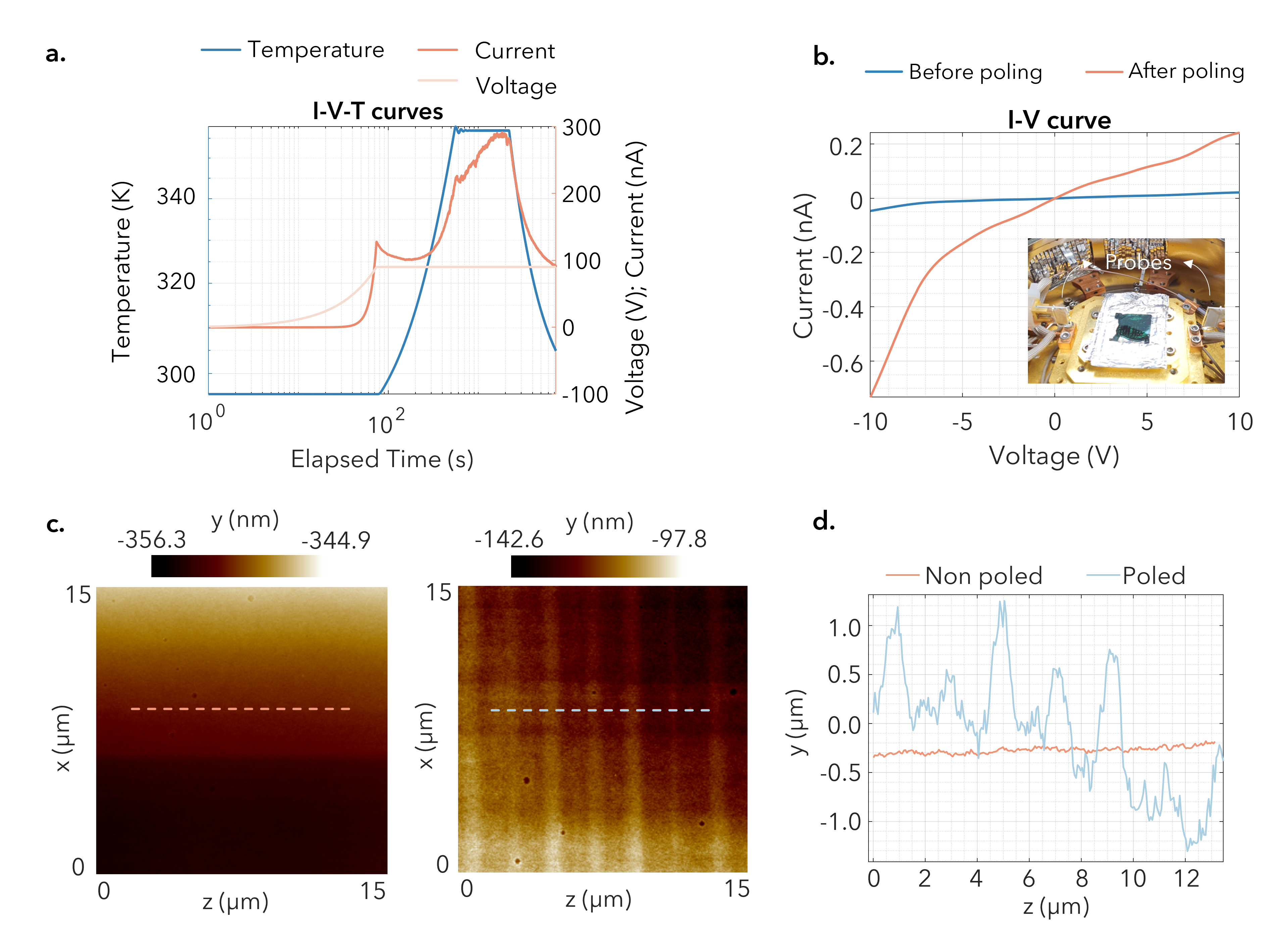}
    \caption{\textbf{Nonlinear activation}. \textbf{a.} The IVT curves of the poling. During the poling and after reaching the glass transition temperature, the current conducted by the polymer between the two electrodes reaches $\approx\;300$nA. After tuning off the heater, the current starts to go down. \textbf{b.} DC measurement before and after poling. An increase of current is observed after the poling. \textbf{c. $\&$ d.} AFM measurements of the poled and unpoled devices. A change in the topography of the polymer is observed after the poling. Red and blue cut lines show the height of the not-poled and poled devices, respectively.
    }
    \label{s-fig:poling}
\end{figure}

To perform the poling procedure, first, the chip is placed in nitrogen atmosphere and is electrically connected to DC probes in a temperature-controlled environment. The nitrogen purge stays constant during poling. After, we apply gradually increasing electric field up to 100 V $\mu m^{-1}$. After the applied electric field reached  100 V$\mu m^{-1}$, we start to increase the temperature  to a poling temperature of $T = 84^\circ C$ at a rate of $\approx 20^\circ\text{C}\text{min}^{-1}$. The poling temperature should be optimized for different substrates and nano-photonic architectures, since they have different thermal conductivities and heat distributions during poling, which could require adjustments to the poling schedules. After reaching the required poling field and poling temperature, poling takes place. To ensure homogeneous and complete poling of the molecules, the chip is held at the temperature $T=84^\circ\text{C}$ for 5-10 minutes. A current rise may be observed as a sign of poling. At the end, the heater is turned off,  and the chip starts to cool gradually to near room temperature. The poling field should be on until the sample reaches at least $40^\circ\text{C}$. Fig. \ref{s-fig:poling} a-b shows the IV and IVT measurements before and after poling. For freshly spin-coated samples, an increase in the current may be observed after poling, which is an indication of poling. Another indication of the poling is the change in the topography of the polymer. One way to verify this is by measuring the height of the film by Atomic Force Microscope (AFM). For our geometry, poling leads to a periodic topography that is aligned with the interdigitated array of electrodes, as visualized in Fig.~\ref{s-fig:poling} c-d (two-dimensional map and cut-lines, respectively).

\section{Experimental setup}\label{supp:Experimental setup}
\subsection{Reflection and transmission characterization setup}\label{Supp:sub:Reflection and transmission characterization setup}
The resonance properties of the electro-optic metasurface were first characterized in a reflection/transmission measurement setup, shown in Fig. \ref{s-fig:tls}a. A tunable laser source (TLS, Keysight 8164B) is first linearly polarized to the polarization state of interest via a Half Wave Plate (HWP, Thorlabs	
WPHSM05-1550) and a Linear Polarizer (LP, Thorlabs LPNIR050). The mirrors M1 and M2 are used to align the light to normal incidence with respect to the device. Prior to the metasurface, the incident light of $2w_0=400 ~\mu$m is focused onto the device by a first lens (L1) with a focal length of $f
_1=10~$cm. The transmitted beam is collimated back by a second lens (L2) with a focal length $f
_2=20~$cm and detected by a photodiode (PD1, Nirvana Auto-Balanced Optical Receiver, Model 2017). The reflected beam from the chip is collimated by the first lens (L1), reflects back from the beamsplitter and is collected by the same photodiode. The red line represents the transmitted light, and the green dashed line represents the reflected light. The measured reflected powers are then normalized to the references to obtain the reflectances that are shown in Fig. \ref{s-fig:tls}b-i for device I with the pitch size of $1.04\;\mu m$ and device II with the pitch size of $1.06\;\mu m$. For the transmission reference, the device is removed, and the transmission is measured with the same components. For the reflectance reference, the device is replaced with a gold mirror (Thorlabs, PF10-03-M01), and the measured reflectance is considered as the reference. 

An important property of our metasurfaces is their modulation efficiency, which quantifies the amount of intensity modulation provided by the metasurface in reflection upon an applied bias $V_m$. The modulation efficiency of the devices, defined as $\displaystyle\left.\eta(\lambda)\right|_{V_m} = \frac{dI_R(\lambda)}{I_{\mathrm{max}}}|_{V_m}$, is then characterized with the same setup, to also confirm the nonlinear activation of the devices. $I_{max}$ is the reflected intensity at the maximum of the reflectance. For this purpose, the output of PD1 and PD2 is connected to a lock-in amplifier (Zurich Instruments UHFLI), allowing characterisation in both transmission and reflection. An Arbitrary Wave Generator (Zurich Instruments HDAWG) applies the RF modulation of frequency $\nu_{mod}$ to the device and sends the signal with the same frequency as the reference to the lock-in amplifier. The TLS sweeps the wavelength and the modulation of the reflected/transmitted light is measured with the lock-in amplifier for each wavelength. Fig.~\ref{s-fig:tls}b-ii shows the measurement results of the modulation characterization for devices I and II. The modulation is greatest where the slope of the reflectance is maximal. 

\subsection{Frequency and amplitude modulation setup}\label{Supp:sub:Frequency and amplitude modulation setup}

\begin{figure}
    \centering
    \includegraphics[width=0.7\linewidth]{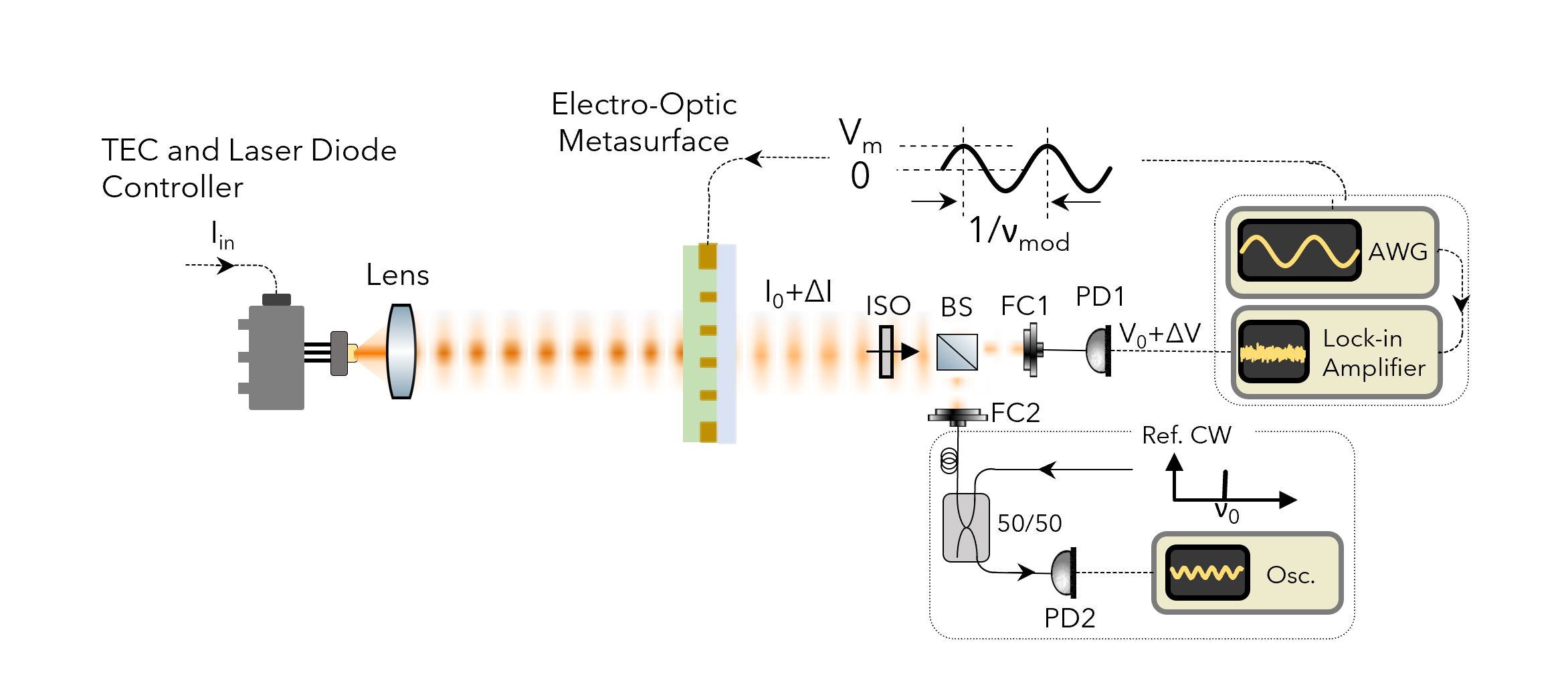}
    \caption{\textbf{Measurement Setup}. 
     The output of the laser diode is modulated by the electro-optic metasurface. The transmitted beam is split by a beam 
    splitter into two paths: one directed to the homodyne detection setup for amplitude modulation measurements, and the other 
    to the heterodyne detection configuration for frequency tuning measurements.
    FC: fiber coupler; BS: beam splitter; CW: continuous wave; AWG: arbitrary waveform generator; PD: photodiode, Osc.: oscilloscope.
    }
    \label{s-fig:HD_eo_setup}
\end{figure}

\begin{figure}
    \centering
    \includegraphics[width=1.0\linewidth]{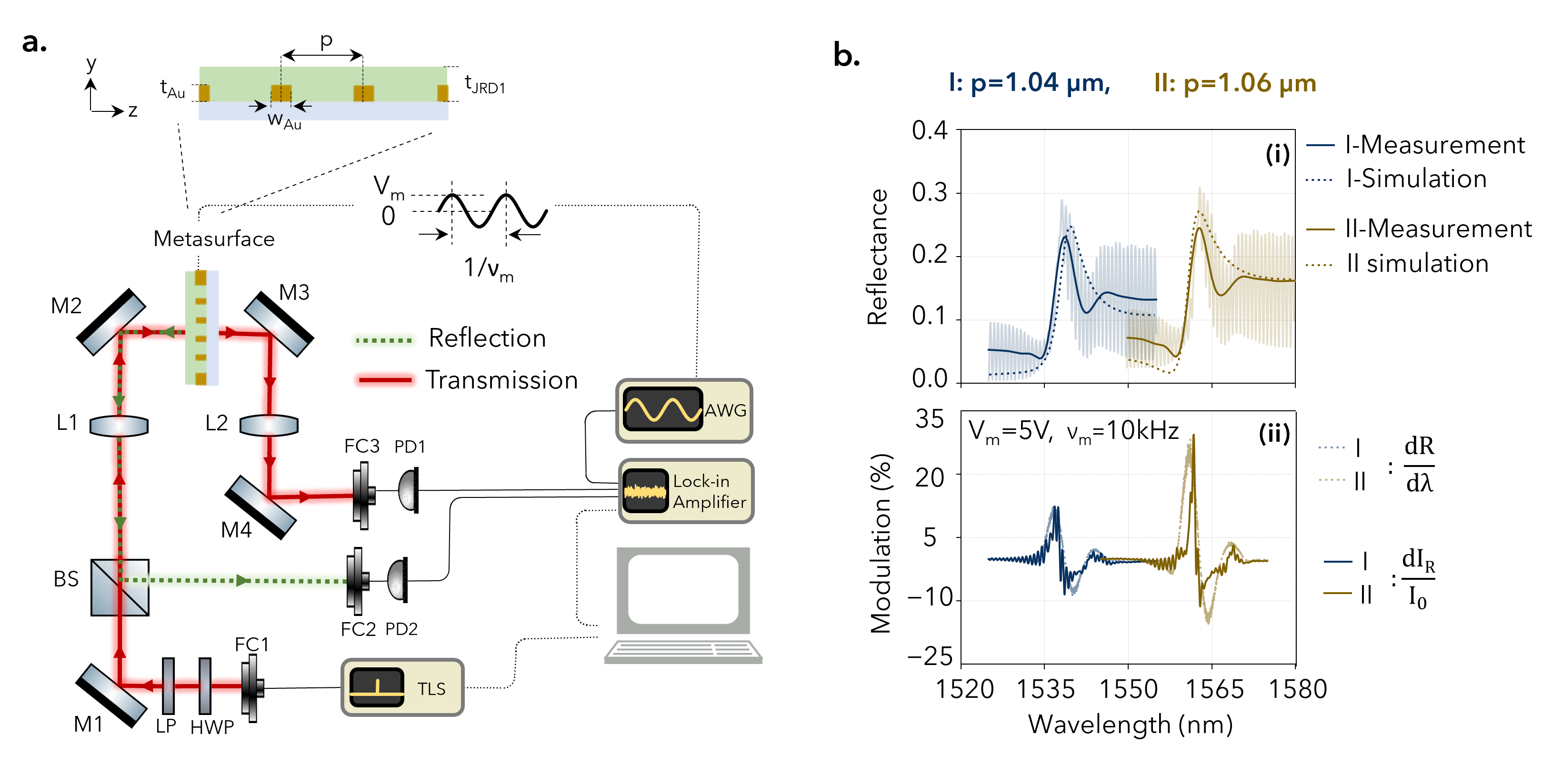}
    \caption{\textbf{Device characterization.} 
\textbf{a.} Schematic of the experimental setup. Red lines indicate the transmission path, and green lines represent the reflection path. The metasurface consists of a periodic array of gold electrodes with pitch size $p$, thickness $t_\mathrm{Au} = 50~\mathrm{nm}$, and width $w_\mathrm{Au} = 200~\mathrm{nm}$, deposited on a nonlinear material with thickness $t_\mathrm{JRD1} = 1.8~\mathrm{\mu m}$.
\textbf{b.} (i) Measured reflectance spectra of two devices with pitch sizes $p = 1.04~\mu\mathrm{m}$ (Device I, blue) and $p = 1.06~\mu\mathrm{m}$ (Device II, gold). Fabry–Pérot fringes are retained (shown transparently) for spectral accuracy. (ii) Modulation efficiency measured by the lock-in amplifier as a function of wavelength under sinusoidal RF driving ($V_m = 5~\mathrm{V}$, $\nu_{mod} = 10~\mathrm{kHz}$). The normalized modulation depth $dI_R / I_{max}$ (where $I_{max}$ is the peak reflected intensity) follows the derivative of the measured reflectance $dR/d\lambda$ (dotted lines), confirming maximal modulation efficiency happening when the slope of the reflectance is maximal. Simulated reflectance spectra (dotted lines in (i)) are obtained using CST unit-cell modeling. The substrate is modeled as fixed silica, and the refractive index of JRD1 is implemented using a dispersive model provided by the manufacturer. A chromophore concentration of 50\% is assumed, matching experimental conditions. TLS: Tunable Laser Source; L1, L2: Lenses; M1–M4: Mirrors; BS: Beam Splitter; LP: Linear Polarizer; HWP: Half-Wave Plate; FC: Fiber Coupler.
}
    \label{s-fig:tls}
\end{figure}

The setup used for measuring the frequency and amplitude modulation introduced by the metasurface is shown in Fig.~\ref{s-fig:HD_eo_setup}.

An anti-reflection-coated laser diode (SAL-1550-010) is used as a coherent light source, controlled by the ITC4001 TEC module, which provides the required injection current and maintains a constant temperature of the laser diode. An isolator is placed after the electro-optic metasurface to prevent any parasitic reflections from returning to the laser diode.  To achieve optimal alignment of the metasurface with the gain medium, an optical spectrum analyzer (Anritsu MS9740B) is used to measure the emission spectrum below threshold, where the resonance of the metasurface is visible in transmission. For evaluating the sideband suppression when operating single mode we use a high-resolution heterodyne optical spectrum analyzer (Apex AP2043B).

For measuring the intensity modulation, a balanced photodiode (PD1, Thorlabs PDB480C-AC) with a 1.6~GHz bandwidth collects the light for the homodyne lock-in path.  An arbitrary waveform generator (HDAWG from Zurich Instuments) was used to modulate the device and to provide the reference signal for the lock-in amplifier. The RF output of the photodiode is connected to the lock-in amplifier (UHFLI from Zurich Instruments), while the DC output is connected to a Source Meter (Keithley 2400). All LI curves presented throughout this work were measured using the DC output of PD1 and read via the Keithley. The coupling efficiency of FC1, the splitting ratio of the beam splitter, and the transimpedance gain of the photodiode are all taken into account to calculate the absolute power after the external cavity.

For measuring the instantaneous frequency, we set up a real-time heterodyne measurement, where a continuous-wave laser (Keysight N7776C) beats with the output of the external cavity laser on a fast photodiode with 5~GHz bandwidth (PD2, Thorlabs DET08CFC). The CW laser is set to a frequency close to the lasing frequency (within a few hundreds of megahertz). A high-bandwidth oscilloscope (Keysight Extra604A) records the output of PD2, and the time trace on the oscilloscope is used as the raw data to calculate the instantaneous frequency as a function of time. In each spectrogram (e.g. shown in Fig.~3 of the main manuscript), the central frequency of the heterodyne signal is given by the wavelength difference between the reference CW laser and the lasing frequency of the device.

\section{Metasurface design}\label{Supp:simulations-cst-sweeps}
In this section, we investigate how variations in the geometrical properties of the metasurface influence its resonant behavior. To this end, we perform numerical simulations using the unit cell module of CST Microwave Studio. 
To accurately model the nonlinear response, the dispersive refractive index of a 50\% JRD1:PMMA mixture was implemented in CST. 
The electrode pitch $p$ and the nonlinear material thickness $t_\mathrm{JRD1}$ were used as sweep parameters, 
while the gold width ($w_\mathrm{Au}=200~\mathrm{nm}$) and thickness $t_{Au}=50~\mathrm{nm}$, also thickness of the nonlinear material $t_{jrd1}=1.8~\mu\mathrm{m}$ were kept fixed, matching the fabricated devices.  

\subsection{Optical resonance}\label{supp:sub:CST_psweep}
In our work, we opt to operate our external cavity on the mode shown in Fig.~\ref{s-fig:cst_p}a, which is the first excited guided mode, as visible from the electric field distributions of $E_y$, $E_z$. Sweeping the pitch size between $p=1.01{-}1.07~\mu\mathrm{m}$ allows to shift the reflectance peak across the telecom band, see $r_\mathrm{ext}$ in Fig.~\ref{s-fig:cst_p}b.  Typical maximal reflection coefficients are around $r_{max}=0.25$, and the quality factor of these resonances varies between 100 and 250. Finally, around resonance a phase excursion larger than $\pi$ is observed. The device used for our external cavity experiments discussed in the main text is shown with a thicker line.

\begin{figure}
    \centering
    \includegraphics[width=1.0\linewidth]{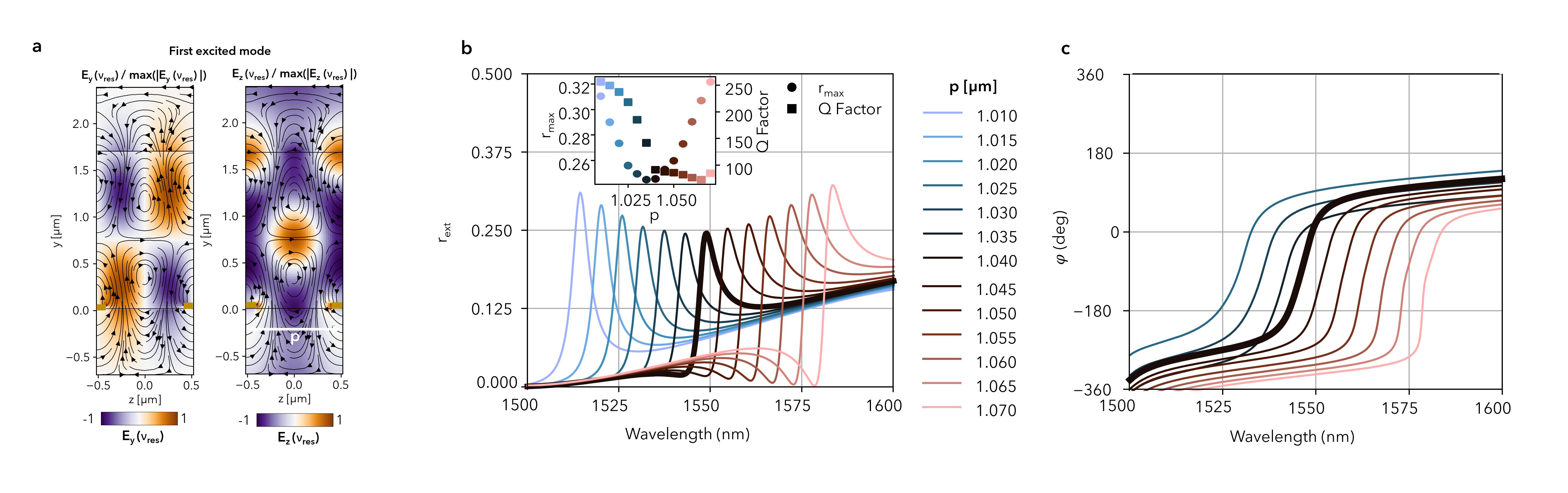}
    \caption{\textbf{Impact of electrode pitch size on resonance properties.}
    \textbf{a}. Simulated field distributions ($E_y$, $E_x$) at resonance for $p=1.02~\mu\mathrm{m}$ and $t_\mathrm{JRD1}=1.7~\mu\mathrm{m}$, 
    showing strong confinement within the nonlinear material.  
    \textbf{b}. Reflectance spectra for varying pitch sizes ($p=1.01{-}1.07~\mu\mathrm{m}$); increasing $p$ leads to a red-shifted resonance and reduced maximum reflectance. 
    Inset: extracted $Q$-factor and maximum reflectance values.  
    \textbf{c}. Phase response of the reflection, centered around zero; larger pitch results in slightly greater phase excursion, enabling stronger phase modulation.  Thick line represents the device used in the experiments of the main text. 
    \label{s-fig:cst_p}
    }
    \label{s-fig:cst_p}
\end{figure}

\subsection{Overlap calculation}\label{supp:sub:ovlcal}

The nonlinear polarization in the orientationally ordered materials, such as polymer JRD1 that we are using in this research, can be described as \cite{singer1987second}

\[
\begin{bmatrix}
P_x(\nu_{opt} \pm \nu_{\text{mod}}) \\
P_y(\nu_{opt} \pm \nu_{\text{mod}}) \\
P_z(\nu_{opt} \pm \nu_{\text{mod}})
\end{bmatrix}
=
\begin{bmatrix}
0 & 0 & 0 & 0 & d_{15} & 0 \\
0 & 0 & 0 & 0 & 0 & 0 \\
d_{31} & 0 & d_{33} & 0 & 0 & 0
\end{bmatrix}
\begin{bmatrix}
2E_x(\nu) E_x(\nu_{\text{mod}}) \\
2E_y(\nu) E_y(\nu_{\text{mod}}) \\
2E_z(\nu) E_z(\nu_{\text{mod}}) \\
E_y(\nu) E_z(\nu_{\text{mod}}) + E_z(\nu) E_y(\nu_{\text{mod}}) \\
E_x(\nu) E_z(\nu_{\text{mod}}) + E_z(\nu) E_x(\nu_{\text{mod}}) \\
E_x(\nu) E_y(\nu_{\text{mod}}) + E_y(\nu) E_x(\nu_{\text{mod}})
\end{bmatrix}.
\]

Here, $P_x$, $P_y$, and $P_z$ are the components of the induced nonlinear polarization at the frequencies $\nu_0 \pm \nu_{\text{mod}}$, and $E_x$, $E_y$, and $E_z$ are the corresponding components of the electric field at the optical frequency $\nu$ and RF $\nu_{\text{mod}}$, interacting in the polymer. The matrix elements $d_{15}$, $d_{31}$, and $d_{33}$ represent the nonzero components of the second-order nonlinear susceptibility tensor $d_{ijk}=-\frac{n^4}{4}r_{ijk}$. 

To analyze the behavior of the optical and RF fields, one needs to solve the nonlinear wave equations, keeping in mind that the nonlinear coefficient $d_{33}(\mathbf{r})$ is space-dependent in our case, since poling is applied via the interdigitated array:
\begin{equation}
\left[ \nabla^2 - \frac{\epsilon_r}{c_0^2} \frac{\partial^2}{\partial t^2} \right] 
E_z^{\text{nl}}(\mathbf{r},t) = \mu_0 \frac{\partial^2}{\partial t^2} 
\big( d_{33}(\mathbf{r}) E_z^{\text{opt}}(\mathbf{r},t) E_z^{\text{mod}}(\mathbf{r},t) \big)
\end{equation}
with the following definitions of the fields
\begin{equation}
E_z^{\text{nl}}(\mathbf{r},t) = E_0^{\text{nl}} \, e_{opt}(z,y) \,
e^{-i(\omega_{\text{nl}} t + \varphi_{\text{nl}}(t))} \,
e^{-i k_y^{\text{nl}} y} \, e^{-i k_z^{\text{nl}} z},
\end{equation}

\begin{equation}
E_z^{\text{opt}}(\mathbf{r},t) = E_0^{\text{opt}} \, e_{\text{opt}}(z,y) \,
e^{-i(\omega_{opt} t + \varphi_{opt}(t))} \,
e^{-i k_y^{opt} y} \, e^{-i k_z^{opt} z},
\end{equation}

\begin{equation}
E_z^{\text{mod}}(\mathbf{r},t) = E_0^{\text{mod}} \, e_{\text{mod}}(z,y) \,
e^{-i(\omega_{\text{mod}} t + \varphi_{\text{mod}}(t))} \,
e^{-i k_y^{\text{mod}} y} \, e^{-i k_z^{\text{mod}} z}.
\end{equation}

\begin{figure}
    \centering
    \includegraphics[width=1\linewidth]{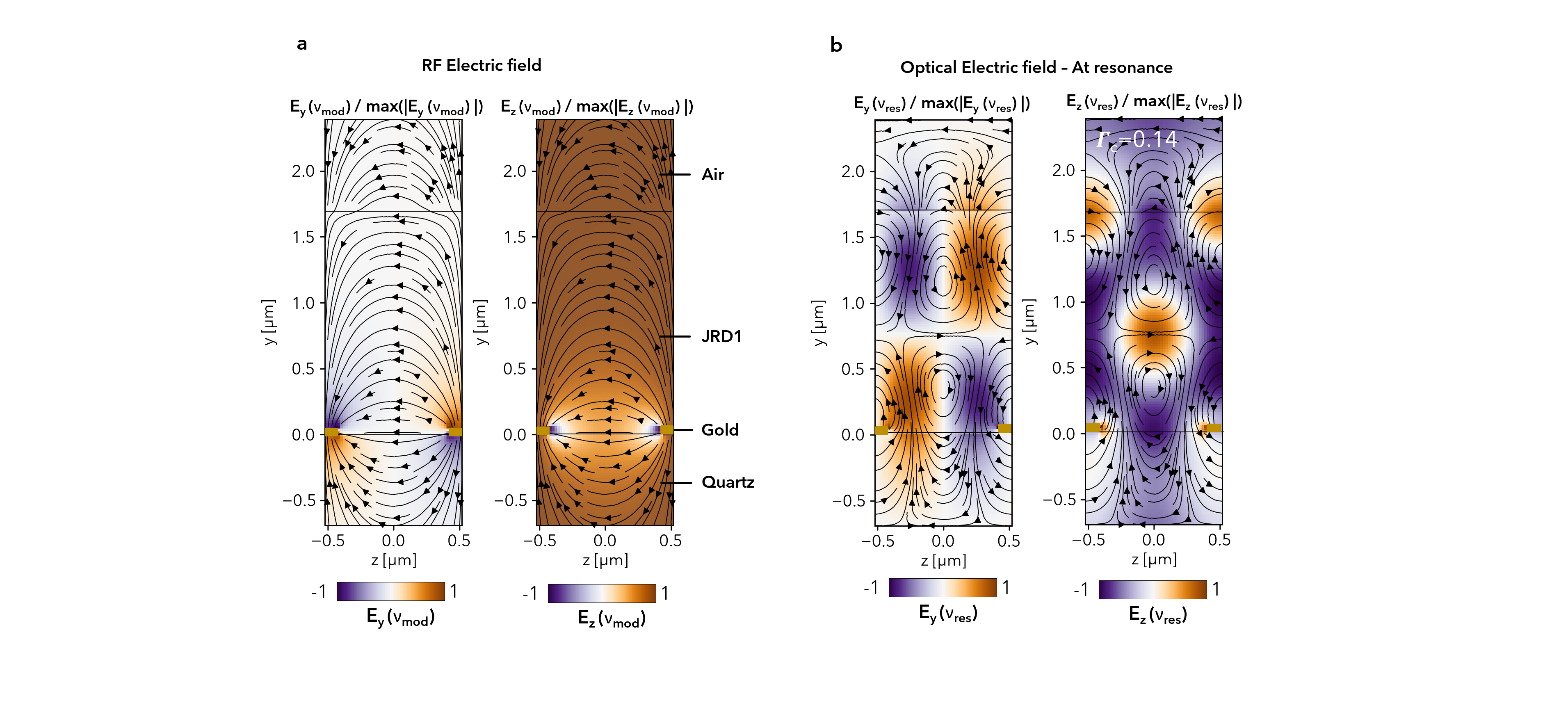}
    \caption{\textbf{Calculation of the spatial overlap factor $\Gamma_c$}. \textbf{a}. RF field applied via the interdigitated array of electrodes. \textbf{b.} Optical field on resonance. Both color plots show the normalized field $E_z$ and $E_y$ and their vectorial orientation in-plane. Here, the optical field, the RF field and the poling fields/$r_{33}$ are in the same plane. The calculated overlap is $\Gamma_c=0.14$.}
    \label{s-fig:ovl}
\end{figure}

Solving this equation as described in \cite{beneaelectro} allows to define the overlap $\Gamma_c$ as

\begin{equation}
\Gamma_c = 
\frac{\displaystyle \iint \big| e_{\text{opt}}(z,y) \big|^2 
       \big| e_{\text{mod}}(z,y) \big|^2 \, dz\,dy}
     {\displaystyle \iint \big| e_{\text{opt}}(z,y) \big|^2 \, dz\,dy}.
\end{equation}

describing the vectorial overlap between the normalized optical field 
$\displaystyle e_{z}^{\text{opt}}(z,y)=\frac{E_{z}^{\text{opt}}(z,y)}{E_0^{\text{opt}}}$ 
and the normalized RF field 
$\displaystyle e_{z}^{\text{mod}}(z,y)=\frac{E_{z}^{\text{mod}}(z,y)}{E_0^{\text{mod}}}$. 
Here, $\displaystyle E_0^{\text{opt/mod}}=
\sqrt{\frac{\iint \lvert E_{z}^{\text{opt/mod}}(z,y)\rvert^2 \, dz\,dy}{S_{\text{eff}}}}$, 
where $\displaystyle S_{\text{eff}}=
\frac{\left(\iint \lvert E_z(z,y)\rvert^2 \, dz\,dy\right)^2}
{\iint \lvert E_z(z,y)\rvert^4 \, dz\,dy}$  is the effective area of the fields. Therefore, $\displaystyle \Gamma_c$ can be calculated directly from the electric fields $E^{opt}(z,y)$ and $E^{mod}(z,y)$ retrieved from numerical simulations. Fig.~\ref{s-fig:ovl} shows the simulation results of the calculated electric field, normalized to $E_0$. The overlap factor obtained for mode this optical mode is $\Gamma_{c}=0.14$. The corresponding RF fields were obtained using the electrostatic solver of CST.

\subsection{Phase modulation across the resonance}\label{supp:sub:phase modulation-sweep}

Achieving a large frequency and amplitude tuning relies on strong changes in the resonance frequency as a function of applied voltage. As discussed in the main text, maximizing these necessitates resonances with large phase slope around the resonance. Since this voltage directly impacts the refractive index of the JRD1 slab, it is insightful to simulate the resonance of the metasurface as a function of the refractive index changes in the nonlinear material and analyze its reflection amplitude and phase as a function of an index shift $\Delta n$ ranging from $-0.004$ to $+0.004$. As shown in Supplementary Fig.~\ref{s-fig:dn}, both the reflection amplitude and the associated phase response remain similar across this range, indicating that the resonance properties are robust to small refractive index variations. From the simulations, a resonance shift of approximately $0.26~\mathrm{THz}$ is observed for a total index change of $\Delta n = 8 \times 10^{-3}$ (Fig.~\ref{s-fig:dn}a and c). By analyzing the phase response, we find its slope at resonance to amount to $\frac{\Delta \varphi}{\Delta \lambda} = 27^\circ$/nm, which is equivalent to $\frac{\Delta \varphi}{\Delta \nu} = 31^\circ$/THz (Fig.~\ref{s-fig:dn}b). 

This phase modulation introduces a frequency modulation of the external cavity laser with a magnitude that depends on the length of the external cavity. In line with formula \ref{eq:lasinglinear} of the main text, shorter cavities enable larger frequency excursion. We model this behavior for varying external cavity lengths and shifts of the resonant frequency $\Delta\lambda_{res}$, at fixed phase slope of $\frac{\Delta \varphi}{\Delta \lambda} = 27^\circ$/nm (Fig.~\ref{s-fig:dn}d). We find that nearly 11 times larger frequency modulation would be possible by decreasing the cavity length to 0.5~mm, which would correspond to attaching the metasurface directly to the facet of the laser diode. 

Finally, the Q-factor of the resonance also plays a major role, since higher Q-factors are associated with steeper phase slopes at resonance. In line with formula \ref{eq:lasinglinear} of the main text, the total frequency excursion depends linearly on this slope. We model this behavior for various shifts of the resonant wavelength in Fig.~\ref{s-fig:dn}e. Simultaneously achieving frequency shifts larger than 1 nm and phase slopes larger than $\frac{\Delta \varphi}{\Delta \lambda} = 70^\circ$/nm would enable frequency excursions larger than 1~GHz, even for an extended cavity length of $L_{ext} = 2.5$~cm.

\begin{figure}
    \centering
    \includegraphics[width=1.0\linewidth]{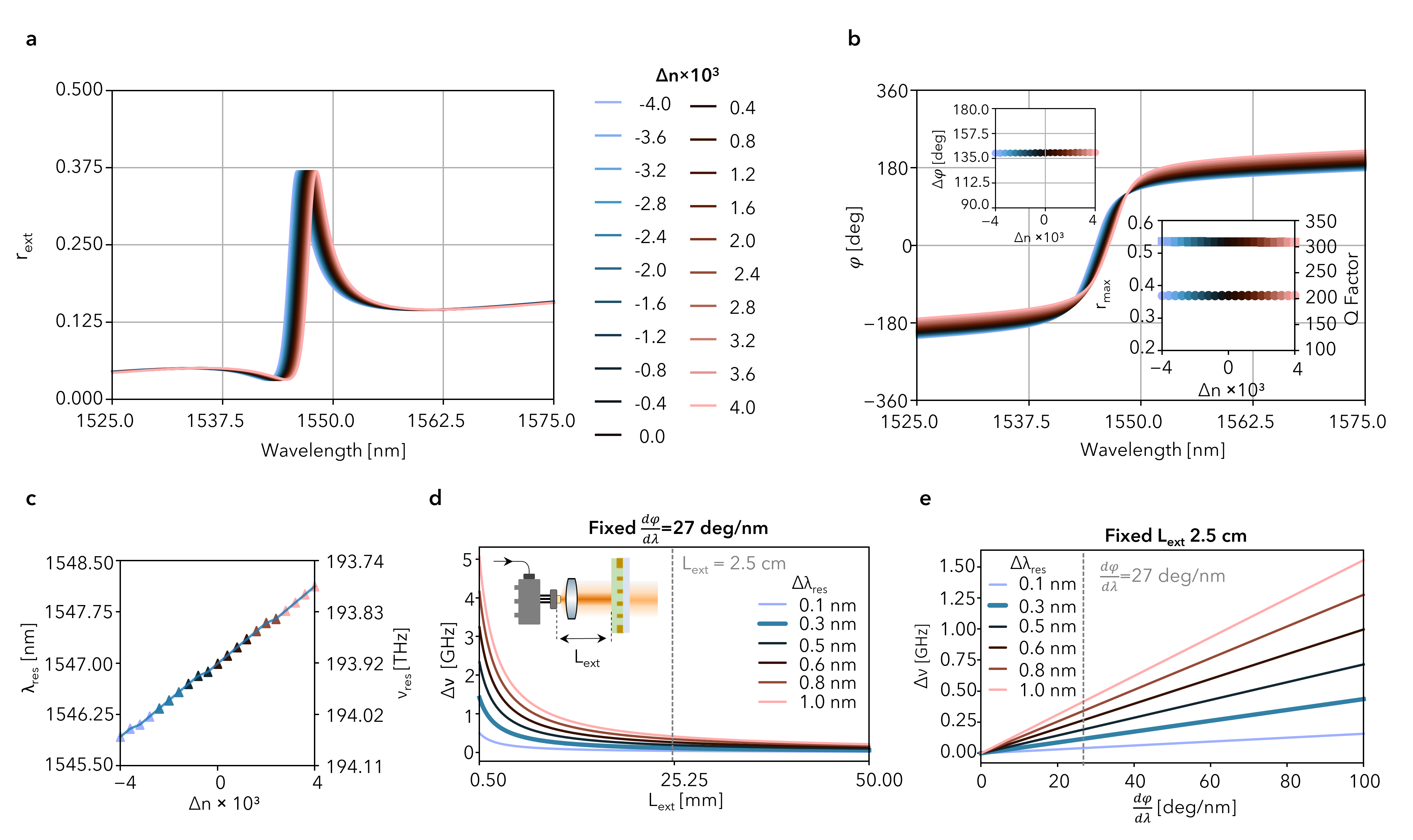}
    \caption{\textbf{Impact of the refractive index shift on the resonance properties}. \textbf{a.} Reflection amplitude and \textbf{b.} Reflection phase for various refractive index changes. The plots are showing the linear simulation of reflection amplitude and phase for various shift of the refractive index of the nonlinear material $\Delta n$, starting from -0.004 to 0.004, where the refractive index of the polymer is defined as $n_\textrm{eo}=n_\textrm{nlm}+\Delta n$.  It can be seen that the resonance properties do not change dramatically across this change. \textbf{c}. Resonance shift due to the refractive index modulation. The shift of $\Delta\nu_\mathrm{res}=$ 0.26 THz ($\Delta\lambda_\mathrm{res}=1.49$ nm) is occurring for a total shift of $\Delta n =8\times 10^{-3}$. Assuming the shift in the resonance frequency with the $\Delta n$ is happening roughly linearly, this shift can be mapped to the measurements. \textbf{d}. Impact of cavity length. For a given electro-optic frequency shift and phase , which is controlled by the applied voltage $V_m$ and the efficiency of the device, and a fixed phase slope, which is detemined by the resonance Q-factor,  one can tune the frequency excursion of the laser $\Delta\nu$ by also changing the external cavity length. Based on the Eq. \ref{eq:lasinglinear} in the main text, the frequency excursion is inversely proportional to the cavity length. Therefore by reducing the cavity length, larger frequency excursions would be achievable. \textbf{e}. Impact of the phase slope (Q-factor) on frequency excursion. For a fixed cavity length and fixed resonance shift, the frequency excursion changes linearly with respect to the phase slope. 
    }
    \label{s-fig:dn}
\end{figure}

\section{Modeling of laser behavior}\label{Supp:sec:Modeling of laser behavior}
The reflection $r_{ext}(\nu)$ directly impacts photon lifetime $\tau_p(\nu)$, saturation photon number $n_{sat}$, and effective absorption cross section $\displaystyle \sigma_{eff}(\nu)=\frac{\sigma_0\cdot g(\nu)}{1+\frac{n_\nu}{n_{sat}}}$. Therefore, changing the properties of the $r_{ext}(\nu)$ directly impacts the number of photons in the cavity at given frequency $n_\nu$ and, as a result, $P_{out}$. To simulate the impact of different resonance properties on the lasing condition, we numerically solve the photon density and population rate equations. We consider the laser a 4 level system, where the lasing occurs between the first and the second levels with $N_1$ and $N_2$ as the lower and upper states populations, and for the sake of simplicity, we neglect the population of $N_0$ and $N_3$ levels \cite{milonni_eberly_lasers_ch3to6,ye2007tunable}.
\begin{align}
\frac{d n_\nu}{dt} &= (N_2 - N_1) \, n_\nu \, \sigma(\nu) c - \frac{n_\nu}{\tau_p} + \sigma(\nu) c N_2 \\
\frac{d N_2}{dt} &= \frac{I_{in}}{e V_c} - (N_2 - N_1) \, n_\nu \, \sigma(\nu) c - \frac{N_2}{\tau_{21}} \\
\frac{d N_1}{dt} &= (N_2 - N_1) \, n_\nu \, \sigma(\nu) c - \frac{N_1}{\tau_{10}} + \frac{N_2}{\tau_{21}}
\end{align}\label{eq:numerical}
The term $I_{\text{in}}$ is the injection current supplied to the laser, $ V_c $ is the active region volume, and $ e $ is the elementary charge. The parameter $ \sigma(\nu) $ stands for the emission cross-section at the lasing frequency which we modeled as a Gaussian function $ \displaystyle\sigma(\nu) = \sigma_0 \exp\left( -\frac{1}{2} \left( \frac{\nu-\nu_{\text{res}}}{\Delta \nu_{gain}} \right)^2 \right) $
where $\displaystyle\nu_{\text{res}}$ is the center frequency at $1550\,\text{nm}$, and $\Delta \nu_{gain} = 75~$nm is the gain bandwidth. $\tau_p $ is the photon lifetime in the cavity, and $\tau_{21} $ and $ \tau_{10} $ are the carrier lifetimes between the respective energy levels. We numerically integrate the coupled rate equations (S6)-(S8) to calculate the unknowns $n_\nu$, $N_1$, and $N_2$. The initial conditions for the photon density and carrier populations are set to
$ [n_{\nu}, N_2, N_1] = [0, 0, 0]. $
The simulation time domain is set to
$ t_{\text{span}} = (0, 5 \, \text{ns}), $
with 10000 time points for evaluation, allowing for the numerical integration of the rate equations under realistic laser cavity and gain conditions, including the impact of wavelength-dependent properties on the lasing condition. Finally, we compute the output power $ \displaystyle P_{\text{out}}(\nu) = \left(1 - r_{ext}(\nu)\right) \frac{\beta_\textrm{sp}n_{\nu} h \nu}{\tau_p(\nu)}$ by solving the steady-state rate equations for each wavelength and injection current and finding steady-state photon number $n_{\nu}$.

We used the following parameters for our calculation: Planck’s constant
$ h = 6.626 \times 10^{-34} \, \text{J}\cdot\text{s}, $
the speed of light
$ c = 3 \times 10^{8} \, \text{m/s}, $ the elementary charge
$ e = 1.602 \times 10^{-19} \, \text{C}, $
the carrier lifetimes
$ \tau_{21} = 0.6 \, \text{ps}$
and  
$ \tau_{10} = 0.1 \, \text{ps}$, the spontaneous emission coupling factor
$ \beta_{\text{sp}} = 10^{-4}, $
the confinement factor
$ \Gamma = 0.3, $
the differential gain coefficient
$ G = 2 \times 10^{-16} \, \text{cm}^2, $
and the background carrier density
$ N_0 = 10^{16} \, \text{cm}^{-3}. $ For the laser cavity parameters, we use a constant reflectivity $r_1=0.99$ and variable $r_{ext}(\nu)$, a cavity length of
$ L_c = 0.5 \, \text{mm}, $
and a cavity width and height of
$ w_c = h_c = 2 \, \mu\text{m}$, yielding a cavity volume of 
$ V_c = w_c \times h_c \times L_c. $ We calculate the mirror (radiative) losses using
$ \displaystyle\alpha_{\text{rad}} (\nu)= -\frac{\ln(r_1 r_{ext}(\nu))}{2L_i}$ and assume the internal losses to be $\alpha_{in} = 7.6~cm^{-1}$.
The total loss coefficient is
$\displaystyle \alpha_{\text{tot}}(\nu) = \alpha_{\text{rad}} (\nu)+ \alpha_{\text{in}}, $ the photon lifetime in the cavity is given by
$\displaystyle \tau_p(\nu) = \frac{1}{\alpha_{\text{tot}}(\nu) c} $ and the
saturation photon number is
$ \displaystyle n_{\text{sat}}(\nu) = \frac{\tau_p(\nu)}{\Gamma \sigma_0 \tau_{21} c}. $

\subsection{Resonance wavelength}\label{supp:sub:modeling_lres-sweep}
 Based on the rate equations, we now study the impact of the resonance wavelength of the metasurface on laser emission. This allows to investigate how the detuning between the peak frequency of the gain $g(\nu)$ and the peak of the reflection $r_{ext}(\nu)$ can impact the threshold current, as well as the emitted spectrum. For the reflection, we assume a Lorentzian-like profile
$ \displaystyle r_{ext}(\nu) = \frac{1}{\delta \nu_{res} - i(\nu - \nu_{res})}, $
where $\displaystyle\nu_{res} = c / \lambda_{res}$ is the resonance frequency of the metasurface and $\displaystyle\delta \nu_{res}$ its linewidth. We assume $\displaystyle\delta\nu_{res}=10\,\text{nm}$. The normalized magnitude of $r_{ext}(\nu)$ was then scaled to match the desired peak reflectivity $r_{max} = 0.4$:
$ \displaystyle r_{ext}(\nu) = r_{max} \frac{|r_{ext}(\nu)|}{\max(|r_{ext}(\nu)|)}. $ We update the corresponding photon lifetime
$\displaystyle\tau_p(\nu) = \frac{1}{\alpha_{\text{tot}}(\nu) c}$, saturation photon number
$\displaystyle n_{\text{sat}}(\nu) = \frac{\tau_p(\nu)}{\Gamma \sigma(\nu) \tau_{21} c}, $ and effective cross-section 
$ \displaystyle\sigma_{\text{eff}}(\nu) = \frac{\sigma(\nu)}{1 + n_{\nu} / n_{\text{sat}}}=\frac{g(\nu)}{(N_2-N_1)}$  at each frequency. 

\begin{figure}
    \centering
    \includegraphics[width=1.0\linewidth]{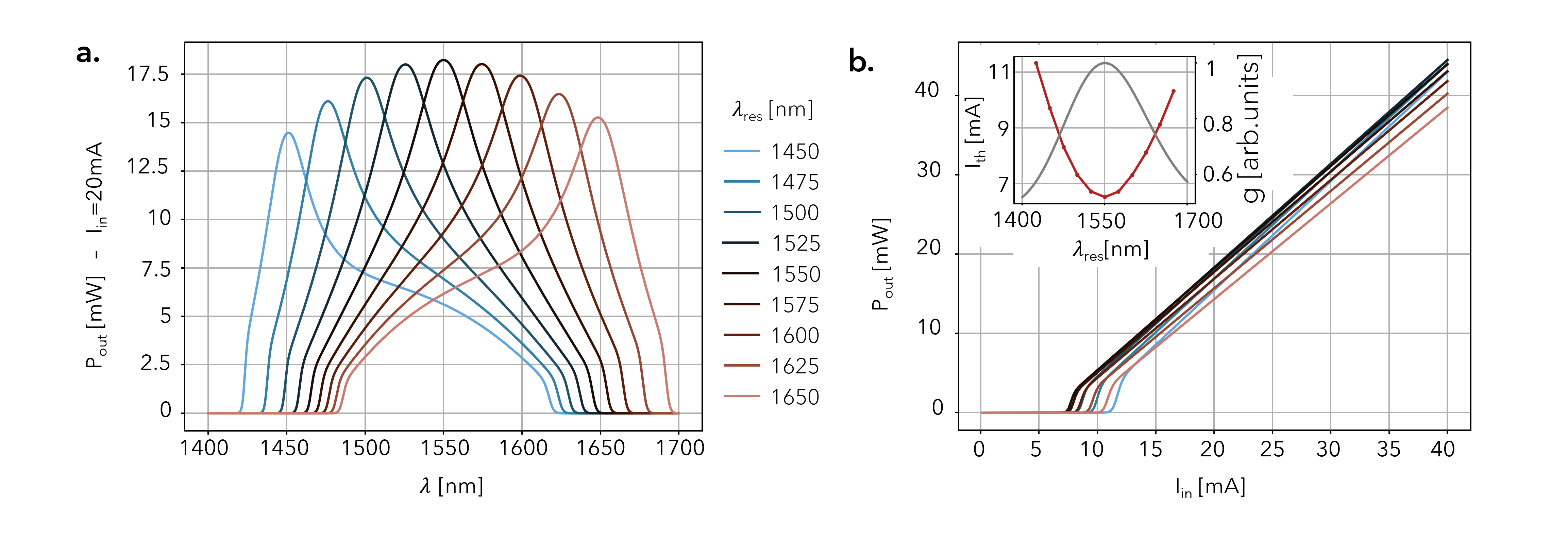}
    \caption{\textbf{Solution of the rate equations for different resonances wavelengths of the external cavity}. \textbf{a}. Output power $P_{out}$ as a function of the de-tuning between the gain peak and the resonance of the metasurface for $I_{in}=20$ mA and $r_{max}=0.4$. The output power is decreasing by increasing the de-tuning. \textbf{b}. PI-curves and threshold current $I_{th}$ for different $\lambda_{res}$. The threshold current is decreasing while decreasing the detuning, and it is minimum for $\lambda_{res}=1550$ nm, where the $g(\nu)$ is maximum.
    }
    \label{s-fig:lres}
\end{figure}

 As a first step, we sweep the values of $\displaystyle\lambda_{res}$ across the wavelength range of 1400~nm to 1700~nm. Fig.~\ref{s-fig:lres}a shows the simulation results of the output power of the external cavity when operated above threshold ($I_{in} = 20~$mA). We observe that the output power decreases with increasing detuning. At the same time, the threshold current also increases (Fig. \ref{s-fig:lres}b and inset, showing relation between threshold current $I_{th}$ and gain $g$ for different $\lambda_{res}$). 

\

This numerical analysis showcases the importance of designing the metasurface such that its resonance aligns with the peak of the gain for achieving low-threshold lasing. 

\subsection{Peak reflection}\label{supp:sub:modeling_visibility-sweep}

\begin{figure}[h!]
    \centering
    \includegraphics[width=0.5\linewidth]{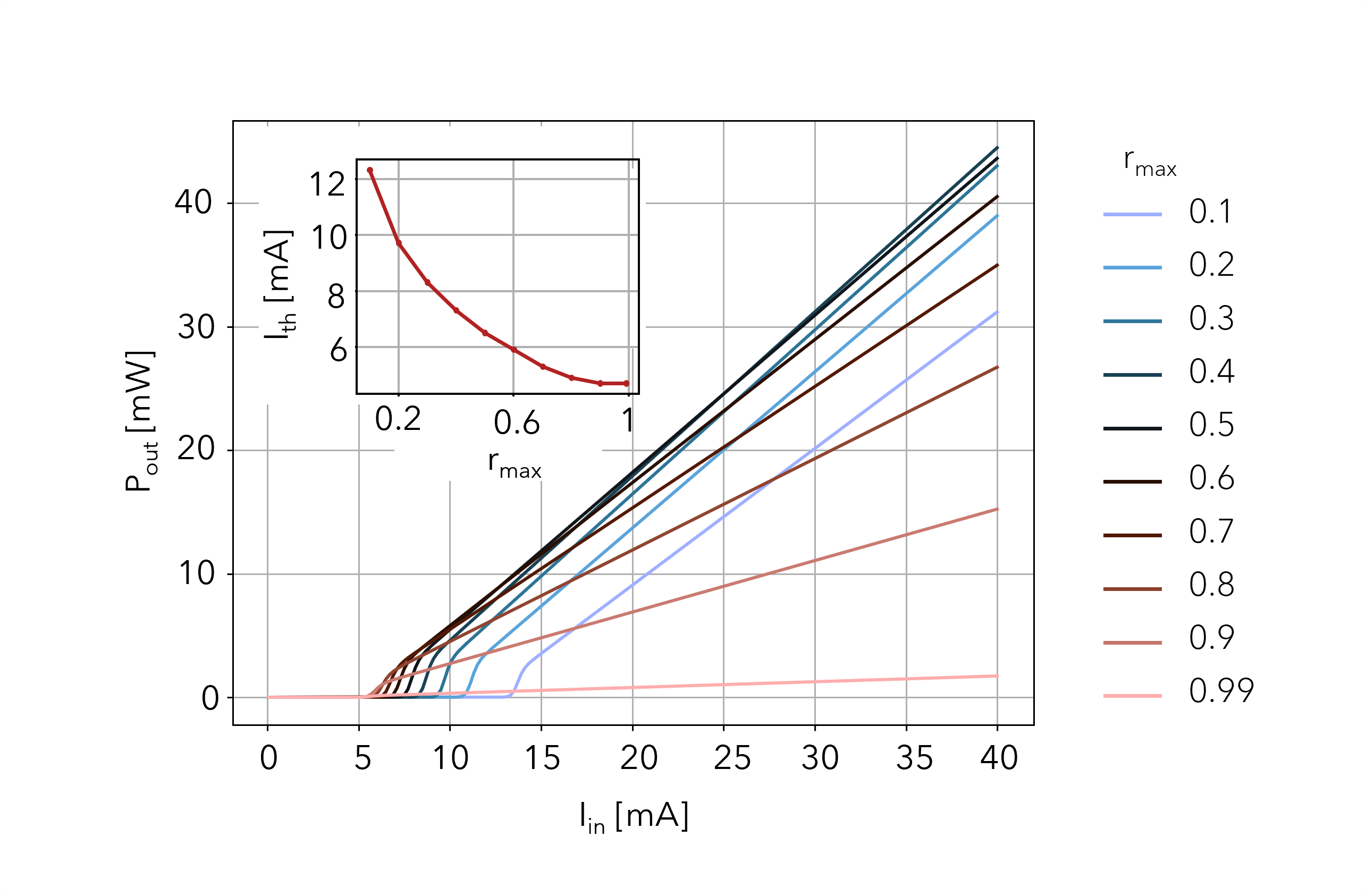}
    \caption{\textbf{Solution of the rate equations for different maximum reflectances of the external cavity.}  LI-curves and threshold current $I_{th}$ for different $r_{max}$. The threshold current is decreasing while increasing the $r_{max}$, and it is minimum for $\lambda_{res}=1550$ nm. The exponential dependence od the $I_{th}$ to $r_{max}$ is highlighted.
    }
    \label{s-fig:rmax}
\end{figure}

The maximum of the reflection $r_{max}$ also impacts the threshold current and the output power of the external cavity. We analyze this behavior by modeling various values of $r_{ext}$ spanning from 0.1 to 0.99, see Fig.~\ref{s-fig:rmax}. As expected, the threshold current is decreasing by increasing the $r_{max}$. At the time, the output power is decreasing when increasing the $r_{max}$, since less light is coupled out of the cavity. We find that our design reflectance of $r_{ext} \approx 0.2$ is a good compromise between the two.

\subsection{Photon number modulation}\label{supp:sub:modeling_modulation}
In the main text we find that above threshold, the metasurface introduces primarily a frequency tuning, since its modulation results in stronger changes in the phase, rather than its amplitude, of the reflection. This aspect is discussed in Note \ref{supp:sub:phase modulation-sweep}. In this section, we complement our analysis by a study where we assume the magnitude of the reflection to be modulated as a function of time as $\displaystyle r_{ext}(t)=r_{max}+\varepsilon\cdot\cos(2\pi\nu_{mod} t)$, where $\epsilon <<1$, and $\displaystyle\nu_{mod}=1$GHz is the modulation frequency. Although we know this modulation to be low (since $\displaystyle\frac{dr}{d\nu}$ is zero at the lasing frequency), we find this analysis to be instructive especially when understanding the increased intensity modulation of the output power around threshold. The output power depends on the intracavity power $\displaystyle P_{in}=\beta h\nu c\frac{n_\nu}{2L_i}$ as $\displaystyle P_{out}=(1-r_{ext}(\nu)^2)P_{in}$ which is related to $r_{ext}(\nu)$. 

\begin{figure}
    \centering
    \includegraphics[width=1.0\linewidth]{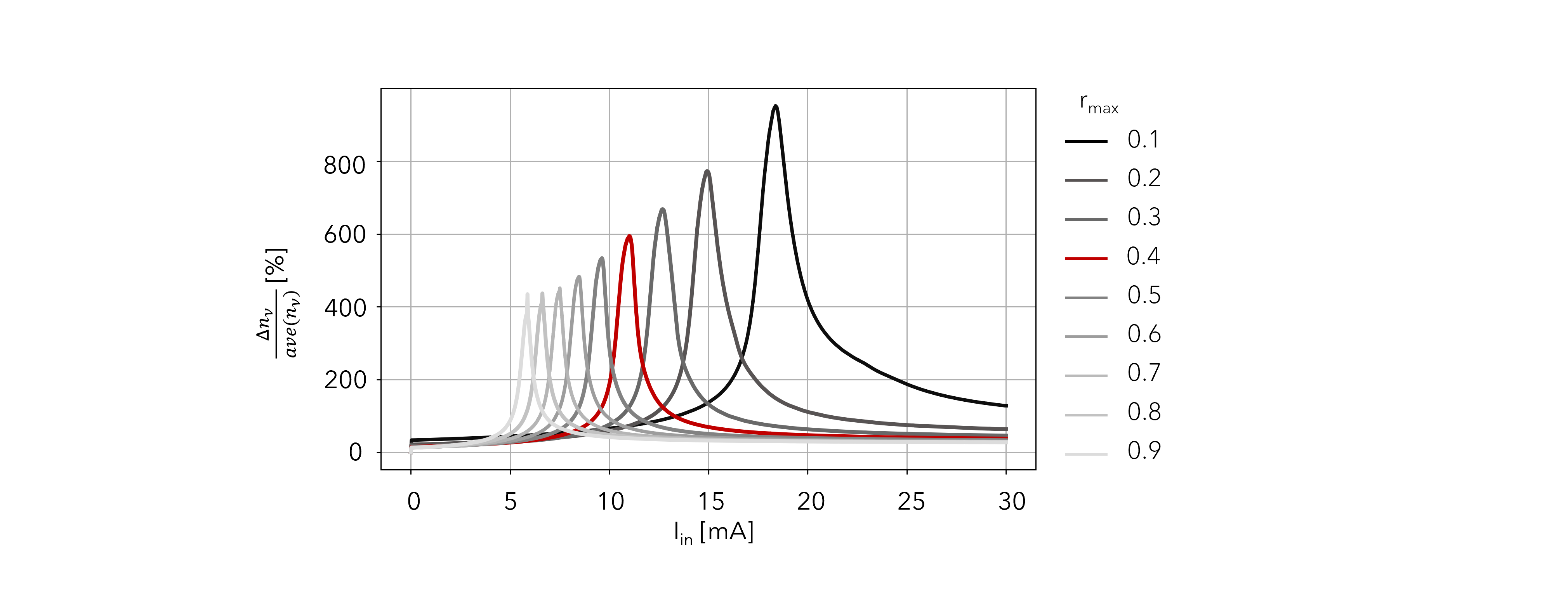}
    \caption{\textbf{Modulation of number of photons in the cavity.}
 The plot shows the modulation of $n_\nu$ for different maximum reflection $r_{max}$. As it was shown before the threshold current decreases with the increase of $r_{max}$. As a result, the injection current at which the modulation is maximum is also reduced by increasing the $r_{max}$.}
    \label{s-fig:modulation1}
\end{figure}

We study the impact of a time-varying $r_{ext}(t)$ on the number of photons in the cavity for fixed $\epsilon = 0.05$ and varying $r_{max}$. We plug the modulated $r_{ext}(t)$ into the rate equations and solve for $n_\nu(t)$. We then calculate the ratio $\displaystyle \eta_{n_\nu}=\frac{\Delta n_\nu}{mean(n_\nu)}$, with $\Delta n_\nu$ being the peak-peak photon number variation. This quantifies the efficiency with which the number of photons in the cavity is modulated. The ratio $\eta_{n_\nu}$ is plotted in Fig. \ref{s-fig:modulation1}a for various $r_{max}$. The ratio is calculated for injection currents from 0 to 30~mA. It can be seen that the modulation efficiency is maximum around the threshold of each device, consolidating the measurement results shown in Fig.~4 of the main manuscript. By increasing the $r_{max}$ the threshold current is decreasing, however, $\eta_{n_\nu}$ is also decreasing. 

\

\begin{figure}
    \centering
    \includegraphics[width=1\linewidth]{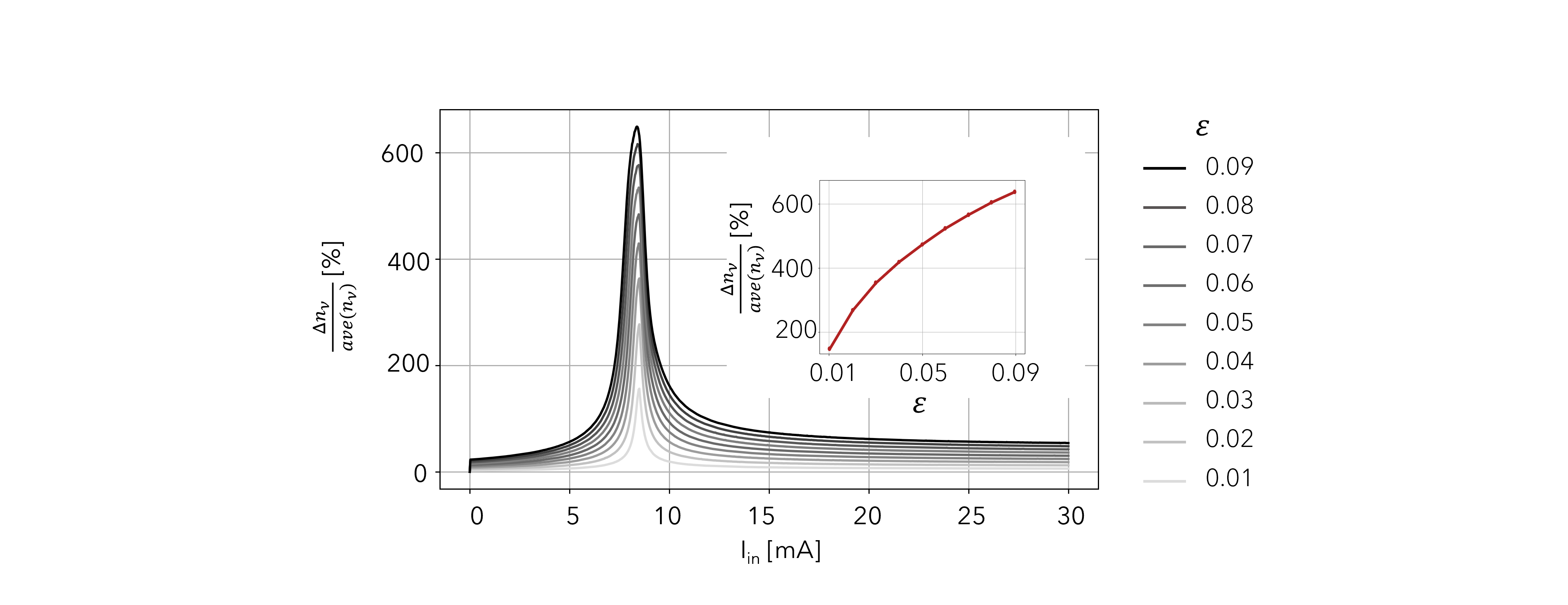}
    \caption{\textbf{Intensity modulation for varying reflectances}.
 $\varepsilon$ is swept from 0.01 to 0.09. The relative change in intra-cavity photon number is then calculated from the rate equations for various injection currents $I_{in}$. The modulation of $n_\nu$ is increasing nonlinearly with $\varepsilon$.}
    \label{s-fig:modulation2}
\end{figure}

The modulation also depends on the variation amplitude $\varepsilon$, which is equivalent to $\Delta R$ induced by the RF. To study the impact of this parameter on the modulation of $n_\nu$, we fix $r_{max}$ at 0.4 and sweep $\varepsilon$ from 0.01 to 0.09 (Fig.~\ref{s-fig:modulation2}). 

\section{Supporting measurements}
\label{Supp:sec:supporting measurements}
\subsection{Spectral shaping}\label{Supp:sub:s shaping}
\begin{figure}
    \centering
    \includegraphics[width=1.0\linewidth]{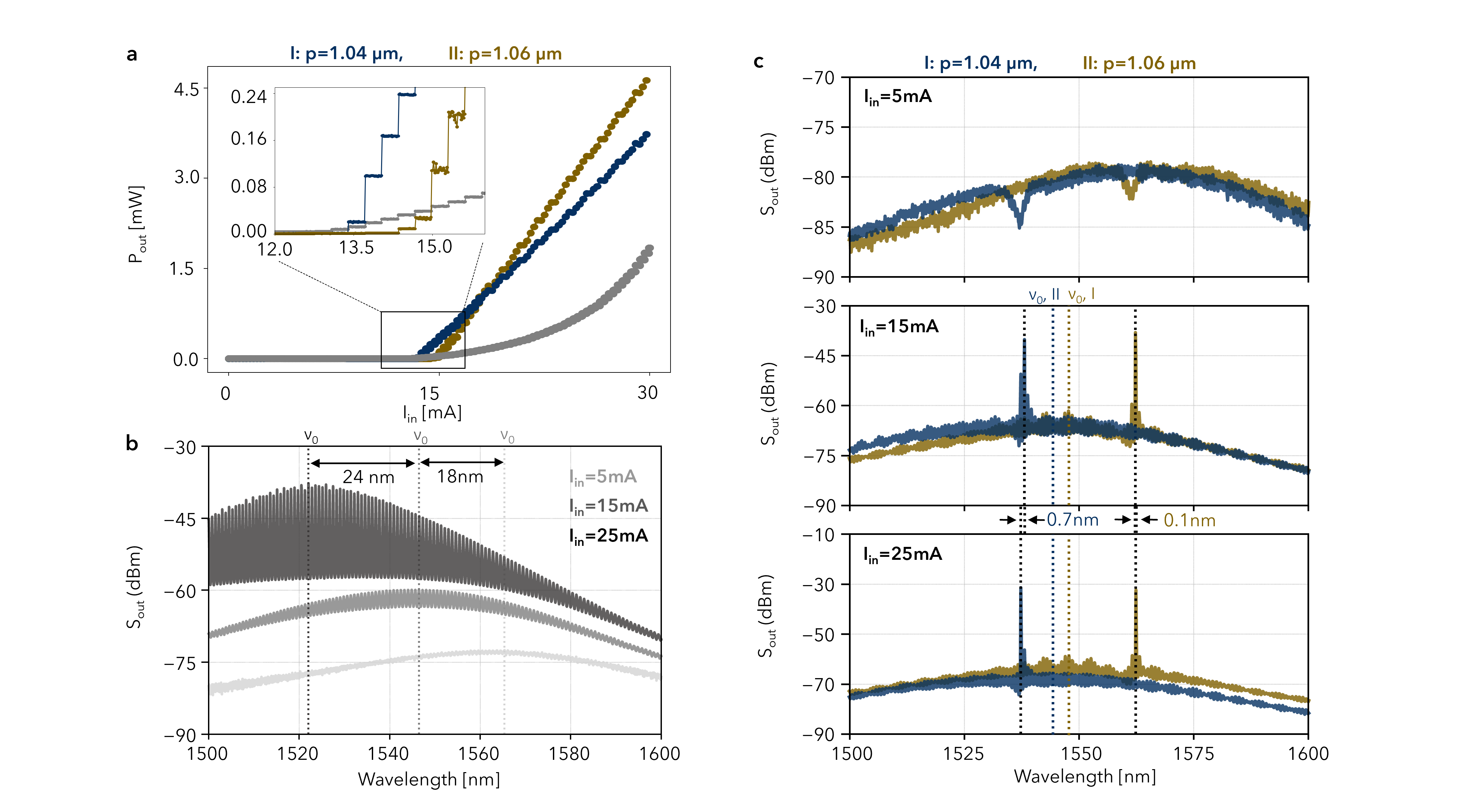}
    \caption{\textbf{Comparison of IV and spectral characteristics for external-cavity devices with different resonance wavelengths}. a. Output power $P_\mathrm{out}$ versus injection current $I_\mathrm{in}$ for Device~I ($p=1.04~\mu\mathrm{m}$, blue) and 
    Device~II ($p=1.06~\mu\mathrm{m}$, gold). 
    Device~II, with higher $r_\mathrm{max}$ but resonance further from $\nu_0$, shows a reduced threshold current. b. Emission spectra of the solitary diode at $I_\mathrm{in}=5$, $15$, and $30~\mathrm{mA}$. 
The spectra exhibit a pronounced blue-shift with current, attributed to the carrier-induced shift of the quasi-Fermi levels. 
The total shift amounts to $\sim42~\mathrm{nm}$ between $5$ and $30~\mathrm{mA}$. c. Emission spectra of the external-cavity devices under feedback for $I_\mathrm{in}=5$, $15$, and $25~\mathrm{mA}$. 
In both devices, the lasing frequency is locked to the external-cavity resonance, and the blue-shift is strongly suppressed. 
The residual shift is limited to $\sim0.7~\mathrm{nm}$ for Device~I and $\sim0.1~\mathrm{nm}$ for Device~II, 
highlighting the stabilization effect of the external cavity.}
    \label{s-fig:iv_spectra}
\end{figure}

Building on the previous section, where we demonstrated how resonance characteristics can be engineered, in this section we provide further measurements showcasing how they affect the lasing condition. We compare the electrical and spectral characteristics of external-cavity lasers when integrated with metasurfaces of varying resonance wavelength by adapting the pitch size $p =1.04$ (device~I) and $1.06~\mu$m (device~II), respectively. Their corresponding reflectances are shown in Fig.~\ref{s-fig:tls}b-i. When comparing these two devices, we find that device~II has a larger threshold and higher slope efficiency (Fig.~\ref{s-fig:iv_spectra}a). This can be explained by the fact that device~I, although having a lower maximum reflectance $r_\mathrm{max}$,  its resonance is less detuned from the maximum of the gain spectrum $\nu_0$, resulting in a lower lasing threshold.  
Device~II, on the other hand, exhibits a higher $r_\mathrm{max}$, which would reduce the threshold current, 
however its larger detuning from the maximum gain results in an overall larger threshold current. At higher injection levels, however, Device~II achieves larger output 
power due to its higher slope efficiency.  We note that the maximum of the gain is dependent on the injection current. This can be observed in the emission spectra of the solitary diode, shown in Fig.~\ref{s-fig:iv_spectra}b. 
As the injection current increases, the emission undergoes a pronounced blue-shift due to the shift of the 
quasi-Fermi levels with carrier density~\cite{bluegan}. 
Between $I_\mathrm{in}=5~\mathrm{mA}$ and $I_\mathrm{in}=30~\mathrm{mA}$, the total spectral shift reaches approximately 
$42~\mathrm{nm}$. Overall, these results highlight the intrinsic trade-off between lasing threshold and output power: a higher $r_\mathrm{max}$ favors lower threshold, while larger detuning increases the threshold. This experimental observation is in line with our modeling (section Note \ref{supp:sub:modeling_visibility-sweep} and \ref{supp:sub:modeling_lres-sweep}). 

In contrast to the solitary diode, the lasing frequency is pinned to the metasurface resonance when operated in the external-cavity configuration. This becomes evident when comparing the spectral shape of the two configurations (Fig.~\ref{s-fig:iv_spectra}c). Furthermore, this frequency locking is robust across injection currents. Our metasurface design featuring one single resonance across the entire gain bandwidth serves to additionally prevent mode hopping. 
For Device~I, the lasing wavelength shifts by only $\sim0.7~\mathrm{nm}$ between $I_\mathrm{in}=15$ and 
$25~\mathrm{mA}$, while for Device~II the residual shift is further reduced to $\sim0.1~\mathrm{nm}$ over the same current range. 
These results demonstrate that the external cavity not only selects the lasing frequency, 
but also stabilizes it against carrier-induced detuning.

\subsection{Side band analysis}\label{Supp:sub:sbr}

An important figure of merit of any single mode laser is the side mode suppression ratio~(SMSR). This ratio describes the amount by which other modes are suppressed compared to the lasing mode. In this section, we analyze the SMSR of our external cavity laser and demonstrate its robustness across injection current. We distinguish two types of SMSR, related to the internal ($\mathrm{SMSR_i}$) and external cavity modes($\mathrm{SMSR_e}$), respectively. These two sets of modes have different free spectral ranges of $\mathrm{FSR_i} = 90~$GHz and  $\mathrm{FSR_e} = 5$~GHz. 

\begin{figure}
    \centering
    \includegraphics[width=1.0\linewidth]{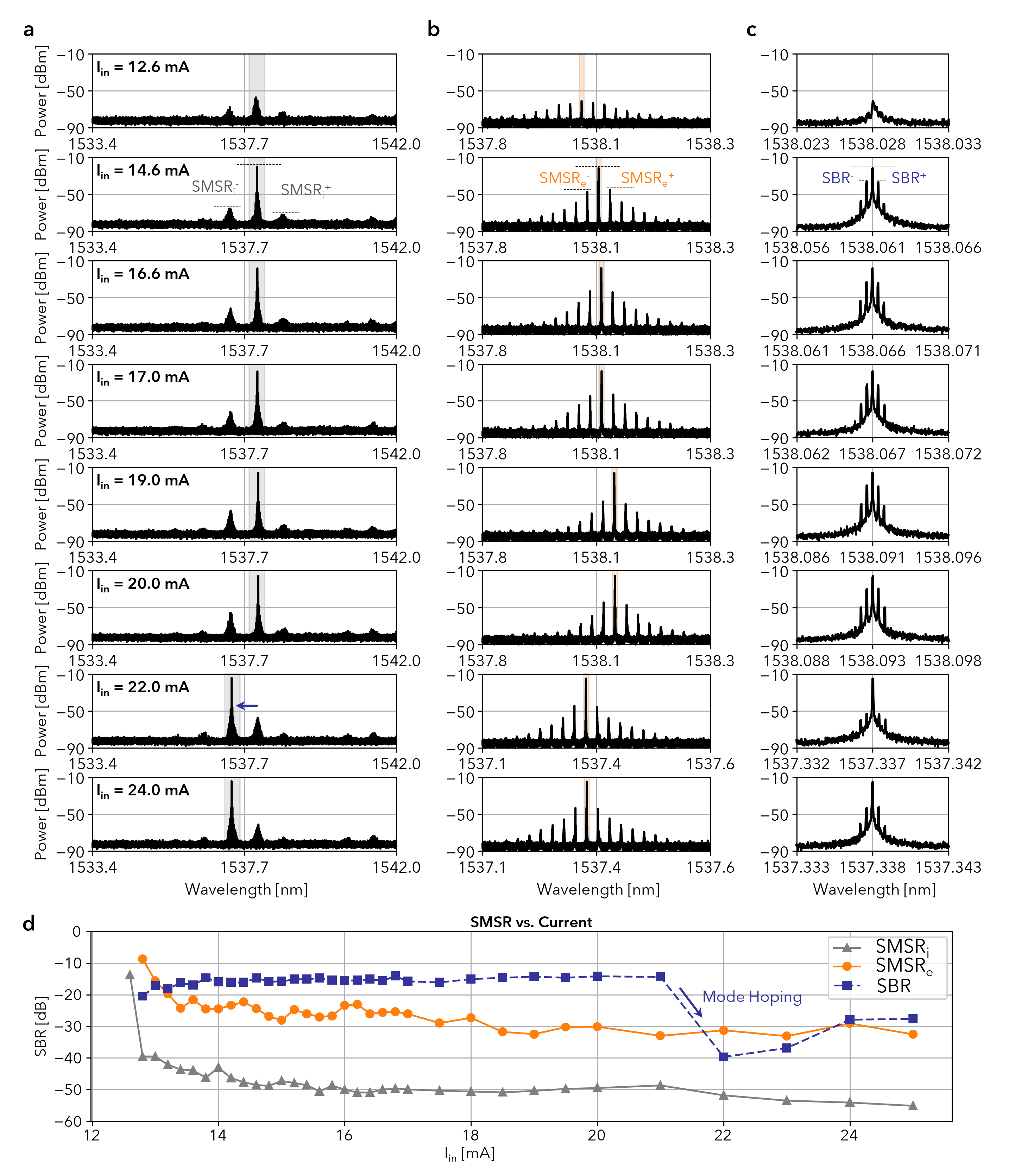}
    \caption{\textbf{Side band ratio analysis}. a.  Optical spectra recorded at different injection currents, including a dataset below threshold at 
    $I_\mathrm{in}=12.6~\mathrm{mA}$ for comparison. Grey shading highlights the external-cavity modes, while the orange 
    zoom-in marks the main peak of the external mode and the RF sidebands around it at $\nu_m = 50~\mathrm{MHz}$.  
    b. Current dependence of the internal sidemode suppression ratio $\mathrm{SMSR}_i$, the external sidemode suppression ratio $\mathrm{SMSR}_e$, 
    and the modulation sideband ratio $\mathrm{SMS}$. $\mathrm{SMSR}_i$ decreases steadily with current, 
    reaching $-55~\mathrm{dB}$ at $I_\mathrm{in}=25~\mathrm{mA}$, indicating strong suppression of internal sidebands. 
    $\mathrm{SMSR}_e$ remains relatively stable around $30{-}35~\mathrm{dB}$, confirming robust suppression of neighboring 
    cavity modes. $\mathrm{SBR}$ stays above $40~\mathrm{dB}$ across most of the operating range but drops 
    abruptly at $I_\mathrm{in}\approx 22~\mathrm{mA}$. The zoomed-in spectra in (a) show that this discontinuity coincides 
    with a mode hop, visible as a sudden change in lasing frequency and sideband strength. After the hop, 
    $\mathrm{SBR}$ partially recovers at higher currents.  
    }
    \label{s-fig:sbr}
\end{figure}

Fig.~\ref{s-fig:sbr}a-b compare the measured spectra at different injection currents, including zoom-ins to visualize the external-cavity modes (highlighted in gray).  
For the sake of comparison, one dataset below threshold at $I_\mathrm{in}=12.6~\mathrm{mA}$ is also included. The current dependence of $\mathrm{SMSR}_i$ is shown in Fig.~\ref{s-fig:sbr}-b. The evolution of $\mathrm{SMSR_i}$ and $\mathrm{SMSR_e}$ as a function of injection current is then visualized in Fig.~\ref{s-fig:sbr}d. Starting from threshold at 
$I_\mathrm{in}\approx 13.6~\mathrm{mA}$, $\mathrm{SMSR}_i$ decreases rapidly and reaches values as low as $-55~\mathrm{dB}$ 
at $I_\mathrm{in}=25~\mathrm{mA}$, reflecting the effective suppression of internal side modes with increasing current.  The external SMSR ratio $\mathrm{SMSR}_e$, on the other hand, also improves with injection current, 
reaching better than $30~\mathrm{dB}$ at $I_\mathrm{in}=25~\mathrm{mA}$. This disparity between the external and internal side mode suppression ratio is related to the free spectral range of these modes. We expect that choosing shorter external cavity lengths would result in improved internal side mode suppression ratio, highlighting possible avenues to improve the performance of our platform.

A second important figure of merit is the amount of modulation provided by the metasurface to the external cavity. As shown in the main manuscript, when the metasurface modulates the intracavity field,  either the instantaneous frequency or the output power
are modulated. In the frequency domain, this manifests as sideband creation around the optical modes of the cavity. To investigate the robustness of this modulation efficiency across currents, we apply a modulation frequency of $\nu_{mod} = 50~\mathrm{MHz}$ and record the sideband ratio $\mathrm{SBR}$ which is the ratio between the RF-generated sideband and the lasing peak (Fig.~\ref{s-fig:sbr}c). When tracking it as a function of injection current (Fig.~\ref{s-fig:sbr}d) we notice a stable response up to 21~mA. At this current, mode hopping occurs between two adjacent internal cavity modes (see
Fig.~\ref{s-fig:sbr}a). After the hop, $\mathrm{SBR}$ partially recovers at higher currents. This effect could be mitigated by either having better anti-reflection coatings or opting for metasurfaces with higher $r_{max}$, which may however compromise output power.

\subsection{Frequency tuning}\label{Supp:sub:rms}

In this section, we provide further data related to the frequency tuning using our platform. 
While Fig.~\ref{fig3} in the main manuscript showed electro-optic frequency tuning for an injection current of 
$I_\mathrm{in}=16~\mathrm{mA}$, here we also investigate the corresponding frequency deviation from an ideal triangular waveform in detail. For each modulation frequency, 
the left column shows the heterodyne spectrogram of the beat signal over multiple modulation periods, and the right column 
plots the corresponding frequency deviation from an ideal triangular waveform. We find that slow-frequency drifts impact the linearity stronger at low modulation frequencies. At higher modulation frequencies (above 250~kHz), the RMS deviations remain below 10~MHz at frequencies.

\begin{figure}
    \centering
    \includegraphics[width=0.65\linewidth]{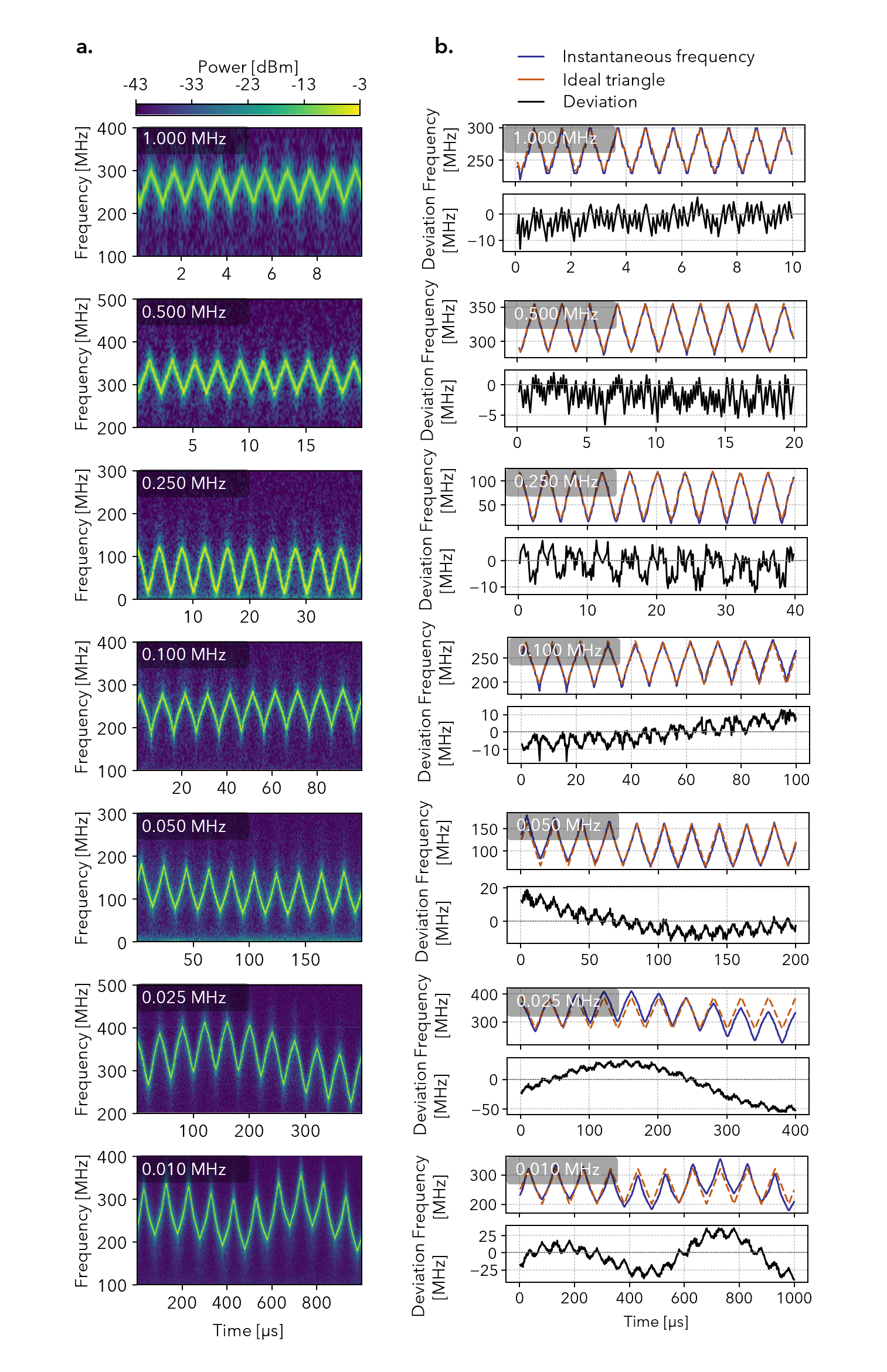}
    \caption{\textbf{Extended electro-optic frequency tuning measurements}. 
     a. Spectrograms (left) and corresponding instantaneous frequency deviations (right) recorded at 
    $I_\mathrm{in}=16~\mathrm{mA}$ for modulation frequencies $\nu_\mathrm{mod}$ ranging from 
    $10~\mathrm{kHz}$ to $1~\mathrm{MHz}$. The excursions show larger RMS values and follow the triangular drive more 
    linearly.  
    b. Same measurements at $I_\mathrm{in}=19~\mathrm{mA}$. Here the excursions exhibit visible nonlinear distortions 
    in the spectrograms and deviation traces, but the corresponding RMS values are lower than at 16 mA.  
    These results extend the main-text measurements by providing the full spectrograms and deviation traces across two bias 
    points and a broad range of modulation frequencies.
    }
    \label{s-fig:eo_spectrograms}
\end{figure}

\listoffigures

\end{document}